\documentclass[preprint,12pt,authoryear]{elsarticle}



\usepackage{natbib}
\usepackage{amsmath}
\usepackage{txfonts}
\usepackage{amssymb}

\usepackage{lipsum} 
\usepackage[T1]{fontenc}
\usepackage{stfloats}
\usepackage{graphicx}
\usepackage[normalem]{ulem}

\usepackage[table,xcdraw]{xcolor}
\usepackage{latexsym,mathrsfs}
\usepackage[hidelinks]{hyperref}
\usepackage{float}
\usepackage{epsfig}
\usepackage{soul}
\usepackage{makecell}
\usepackage{colortbl} 
\usepackage{longtable}
\usepackage{adjustbox}
\usepackage{multirow}
\usepackage{booktabs}
\usepackage{bookmark}
\usepackage{siunitx}
\usepackage{subcaption}

\journal{High Energy Astrophysics}

\begin{document}

\begin{frontmatter}


\title{The effective running Hubble constant in SNe Ia as a marker for the dark energy nature}





\author[1,2,3]{E. Fazzari}{}
\cortext[cor1]{Corresponding author}
\ead{elisa.fazzari@uniroma1.it}

\affiliation[1]{organization={Physics Department, Sapienza University of Rome},
            addressline={P.le A. Moro 5}, 
            city={Rome},
            postcode={00185}, 
            country={Italy}}
            
\affiliation[2]{organization={Istituto Nazionale di Fisica Nucleare (INFN), Sezione di Roma},
            addressline={P.le A. Moro 5}, 
            city={Rome},
            postcode={00185}, 
            country={Italy}}
            
\affiliation[3]{organization={Physics Department, Tor Vergata University of Rome},
            addressline={Via della Ricerca Scientifica 1},
            postcode={00133}, 
            country={Italy}}

\author[4,5,6,7]{M. G. Dainotti}{}
\affiliation[4]{organization={Division of Science, National Astronomical Observatory of Japan},
            addressline={2 Chome-21-1 Osawa, Mitaka}, 
            city={Tokyo},
            postcode={181-8588}, 
            country={Japan}}
            
\affiliation[5]{organization={Astronomy Department, The Graduate University for Advanced Studies, SOKENDAI},
            addressline={Shonankokusaimura, Hayama, Miura District}, 
            city={Kanagawa},
            postcode={240-0115}, 
            country={Japan}}
\affiliation[6]{organization={Space Science Institutes },
            addressline={4765 Walnut St Ste B}, 
            city={Boulder},
            postcode={ 80301}, 
            state={CO},
            country={USA}}
            
\affiliation[7]{organization={Nevada Center for Astrophysics, University of Nevada,89154},
            addressline={4505 Maryland Parkway}, 
            city={Las Vegas},
            postcode={ 80301}, 
            state={NV},
            country={USA}}

\author[8,1]{G. Montani}{}

\affiliation[8]{organization={ENEA, Nuclear Department, C.R. Frascati},
            addressline={Via E. Fermi 45}, 
            city={Frascati},
            postcode={00044}, 
            country={Italy}}

\author[1,2]{A. Melchiorri} {}

\begin{abstract}
We propose a new method that reveal the nature of dark energy (DE) evolution. Specifically, the method consists of studying the evolving trend regarding the effective running Hubble constant: when it increases, it indicates a quintessence nature, and when it decreases, it reveals a phantom behavior.\\
Within the framework of the dark energy models we analyze three parameterizations: the $w$CDM model, a reduced Chevallier-Polarski-Linder (CPL) model and a new theoretical model based on the possible creation of dark energy by the time-varying gravitational field of the expanding Universe.\\
For each DE model, we construct a theoretical effective running Hubble constant, i.e. a function of the redshift, which highlights the difference between modified dynamics and the $\Lambda$CDM-one. Furthermore, these dark energy models are compared to a phenomenological model, called the power-law model (PL), that assumes a decreasing trend of the Hubble constant with redshift, and to the $\Lambda$CDM one. These three theoretical functions for DE are fitted against the binned Type Ia Supernovae (SNe Ia) data samples, i.e. the Pantheon and the Master samples, the latter being a collection of SNe Ia from 4 catalogs: Dark Energy Survey (DES), PantheonPlus, Pantheon and Joint Lightcurve Analysis (JLA), without duplicated SNe Ia, called the Master sample. \\
The main result of our study is that the phenomenological PL model is statistically favored compared to the other proposed scenarios, both for the Pantheon and the Master samples. At this stage, the SNe Ia data do not indicate that the evolution of dark energy models among the studied ones is favored respect to the $\Lambda$CDM. Nevertheless, the binned Pantheon sample allows for a discrimination of the nature of dark energy at least at the $1\,\sigma$ level via the fit of the effective running Hubble constant.
\end{abstract}


\end{frontmatter}
\section{Introduction}
\noindent In recent years, cosmologists have faced an intriguing and puzzling issue: a significant difference, averaging about $5\,\sigma$, between measurements of the Hubble constant from the Planck satellite using Cosmic Microwave Background (CMB) photons \citep{Planck2018}, and from the SH0ES collaboration using Type Ia Supernovae (SNe Ia) as standard candles \citep{SH0ES, scolnic2018, Scolnic_2022, Brout_2022}. This issue is widely known as the ``Hubble tension''. \\
Two main approaches have been proposed to explain this discrepancy: \\
i) an astrophysical bias in data analysis, for instance due to a redshift evolution of SNe Ia, i.e. of the parameters involved in the fundamental standard candle relation \citep{riess1998}; see also similar examples concerning other astrophysical sources in \citet{dainotti-petrosian}; \\
ii) a modification of the cosmological dynamics due to new physics, such as modified gravity, dark energy-dark matter interaction, or exotic and phenomenological modifications of the $\Lambda$CDM-model (for reviews see \citet{whitepaper_cosmoverse, divalentino-Hubbletension}; for specific analyses see \citet{linder2024, montani_hubbletension_1, matcre_montanimary, montani_hubbletension_4, montani_hubbletension_5, Vagnozzi:2019ezj, Vagnozzi:2023nrq, Jiang:2024xnu, Pedrotti:2024kpn}).  
Moreover, values of the Hubble constant different from the central value obtained from SNe Ia are also derived using Gamma-Ray Bursts \citep{dainotti_GRB_23, bargiacchi2023GRB, lenart2023} and Quasars \citep{dainotti2023quasars, dainotti2024quasars} obtained trough new statistical assumptions discussed in \citet{dainotti2024binquasars}.

Recently, particular attention has been devoted to studies showing that the measured value of $H_0$ may depend on the redshift of the sources used \citep{dainotti2021, kazantzidis2020, cinesi_H0z, dainotti_montani2023, Vagnozzi:2023nrq}. For possible interpretations based on fundamental physics, see \citet{schiavone2023, montani-carlevaro-dainotti2024, montani-carlevaro-dainotti2025}.
Specifically, previous analyses \citep{dainotti2021, dainotti2022, dainotti_Master} found evidence of a power-law decrease in the value of $H_0$ across different redshift bins. This result, obtained using both with $\Lambda$CDM-model \citep{weinberg-grav-cosm, Primordial} and the Chevallier-Polarski-Linder (CPL) parametrization for evolutionary dark energy \citep{CPL1, CPL2} to determine $H_0$ in each bin, suggests a possible deviation of the real cosmological dynamics from these models. 

These studies primarily aimed to address the Hubble tension.
However, this paper focuses on a different goal: we explore dynamical models of dark energy (DE) inspired by recent results from the Dark Energy Spectroscopic Instrument (DESI) collaboration \citep{desi, desi2}, which showed a preference for models with evolutionary dark energy over the standard $\Lambda$CDM-model. In particular, DESI findings suggest that the equation of state for dark energy might evolve with time, potentially indicating a transition between quintessence and phantom regimes around a crossing redshift $z_c \sim 0.3$ \citep{extended_desi, Giare_Robust_DDE}. Moreover, as noted in \citet{giare_overviewDDE, giare_dynamical}, intriguing evidence for evolutionary dark energy arises from dataset combinations that exclude both DESI and CMB, such as SDSS-BAO with DESy5 or Union3 SNe Ia. \\
Motivated by these findings, we explore different DE scenarios: the $w$CDM model, a specific case of the reduced CPL parametrization, i.e. the CPL' model ($w_0 = w_a$), as well as a new phenomenological model in which the cosmological gravitational field itself generates dark energy particles as the Universe expands. This latter model modifies the standard cosmological framework described by $\Lambda$CDM. For all three models, we derive a theoretical expression for a redshift-dependent (running) Hubble constant, $\mathcal{H}_0(z)$, and compare our theoretical functions against binned SNe Ia data. The binned results are then compared both with the full SNe Ia samples and with a background-level analysis combining DESI-BAO data, Cosmic Chronometers and the same catalogs of SNe Ia used for the binned analysis. \\
In this context, the objective of our study is to demonstrate how the binned analysis of SNe Ia offers sensitive insights into the dynamic nature of dark energy, particularly its evolving equation of state. 

The manuscript is structured as follows: in Section~\ref{sec2} we define the effective running Hubble constant showing its feature depending from DE nature. Selected theoretical models of DE are presented in Section~\ref{sec3}. The statistical analysis performed and results obtained are shown in Section~\ref{sec4} and~\ref{sec5}. Finally, we discuss the results and provide concluding remarks in Section~\ref{sec6}.

\section{The effective running Hubble constant}\label{sec2}
\noindent Following the definitions provided in \citet{krishnan2021, krishnan2022, schiavone2024}, we introduce the effective running Hubble constant $\mathcal{H}_0(z)$ as a measure of deviation from the standard $\Lambda$CDM model:

\begin{equation}
H(z)=\mathcal{H}_0(z) E(z)^{\Lambda \text{CDM}},
\label{eq:H0_z_1}
\end{equation}
where $H(z)$ is the Hubble parameter of the modified $\Lambda$CDM model, and $E(z)^{\Lambda \text{CDM}}$ denotes the standard $\Lambda$CDM Hubble rate parameter. Consequently, the effective running Hubble constant can be explicitly reformulated as:

\begin{equation}
\mathcal{H}_0(z)= H_0 \frac{E(z)}{E(z)^{\Lambda \text{CDM}}}.
\label{eq:H_0_z_2}
\end{equation}
We emphasize that, since this function measures the discrepancy between modified cosmologies and $\Lambda$CDM dynamics, the effective running Hubble constant reduces exactly to $\mathcal{H}_0(z) \equiv H_0$ in the case of $\Lambda$CDM model. For a similar diagnostic tool of dark energy, see the DESI supporting paper of \citet{extended_desi} where they use $Om(z)\equiv \frac{h^2(z)-1}{(1+z)^3-1}$, with $h(z)=H(z)/H_0$ as defined in \citet{Omz_starobinsky}. In particular, it is a null test of the cosmological constant since the sign of $Om(z)-\Omega_{m0}$ clearly distinguish between cosmological constant (equal to $0$), quintessence ($>0$) or phantom ($<0$) scenarios. \\
It is worth stressing how the theoretical definition of the effective running Hubble constant naturally corresponds to what can be experimentally observed when we detect the value of $H_0$ in different redshift regions. In fact, if the $\Lambda$CDM-model is not properly representing the Universe dynamics, its implementation in data analysis of sources in distant cosmological sites, can not reproduce the same (constant) value of $H_0$, while a given distribution of scattered values has to be recognized. 
It is just this distribution of $H_0$ values that stands for the observational representation of the effective running Hubble constant and it should therefore be fitted via the theoretical expression in eq.\eqref{eq:H_0_z_2}.

Previously, the effective running Hubble constant has been used mainly to explore the Hubble tension \citep{schiavone2022running, montani-carlevaro-dainotti2024, montani-carlevaro-dainotti2025, DeSimone:2024lvy}. Here, we illustrate why and how the effective running Hubble constant also serves as a powerful diagnostic tool to uncover the nature of dark energy. Specifically, we propose a novel approach based on the first derivative of the effective running Hubble constant, $\mathcal{H}_0'(z)$, to determine the nature of dark energy.

We begin by deriving the analytical expression of $\mathcal{H}_0'(z)$ for the $w$CDM model:

\begin{equation}
\begin{aligned}
\mathcal{H}_0'(z) 
&= -\frac{3}{2}\,H_0\,
\frac{1}{(1+z)^{3/2}\,E(z)} \biggl[\,
\frac{(1+z)^5\,\Omega_{m0}\bigl(\Omega_{m0} - (1 - \Omega_{m0})(1+z)^{3w_0}\bigr)}
     {D(z)^{3/2}} \\[1ex]
&\quad + \frac{(1+z)^2\bigl[\Omega_{m0} - (1+w_0)(1 - \Omega_{m0})(1+z)^{3w_0}\bigr]}
     {D(z)^{1/2}}
\biggr].
\end{aligned}
\label{eq:H0prime_theo_wCDM}
\end{equation}
where we introduce 

\begin{equation}
D(z) = \bigl[\,1 + z\bigl(3 + z(3 + z)\bigr)\bigr]\,\Omega_{m0} \,,
\label{eq:Dz}
\end{equation}
and
\begin{equation}
E(z) = \sqrt{\Omega_{m0}(1+z)^3 + (1-\Omega_{m0}) (1+z)^{3(1+w)}} \,,
\label{eq:Ez}
\end{equation}
is the Hubble rate parameter for the $w$CDM model.
\noindent Similarly, for the  CPL model \citep{CPL1,CPL2}, also known as the $w_0w_a\mathrm{CDM}$ parametrization, in which the dark energy equation of state is given by $w(z)=w_0 + w_a \, z/(1+z)$, with $w_0$ and $w_a$ as parameters governing its evolution, the expression becomes:

\begin{equation}
\begin{aligned}
\mathcal{H}_0'(z) 
&= -\frac{3}{2}\,H_0\,
\frac{1}{(1+z)^{3/2}\,E(z)} \biggl[\,
\frac{(1+z)^5\,\Omega_{m0}\bigl(\Omega_{m0} - \Omega_{DE}(z)\bigr)}
     {D(z)^{3/2}} \\[1ex]
&\quad + \frac{(1+z)^2\bigl[\Omega_{DE}(z)\bigl(1 + w_0 + (1 + w_0 + w_a)\,z\bigr) - \Omega_{m0}\bigr]}
     {D(z)^{1/2}}
\biggr].
\end{aligned}
\label{eq:H0prime_theo_w0waCDM}
\end{equation}
where $D(z)$ is the same of above while in this case 

\begin{equation}
\label{eq:Ez_w0wa}
E(z) = \sqrt{\Omega_{m0}(1+z)^3 + \Omega_{DE}(z)} \,,
\end{equation}
with

\begin{equation}
\label{eq:OmegaDE}
\Omega_{DE}(z) = (1 - \Omega_{m0})\,(1+z)^{3\,(1 + w_0 + w_a)}\, 
e^{-3\,w_a\frac{\,z}{1+z}} \,,
\end{equation}
are the Hubble rate parameter and the dark energy density equations for the $w_0w_a\mathrm{CDM}$ model.

\begin{figure}[htbp]
\centering
\includegraphics[width=0.45\textwidth]{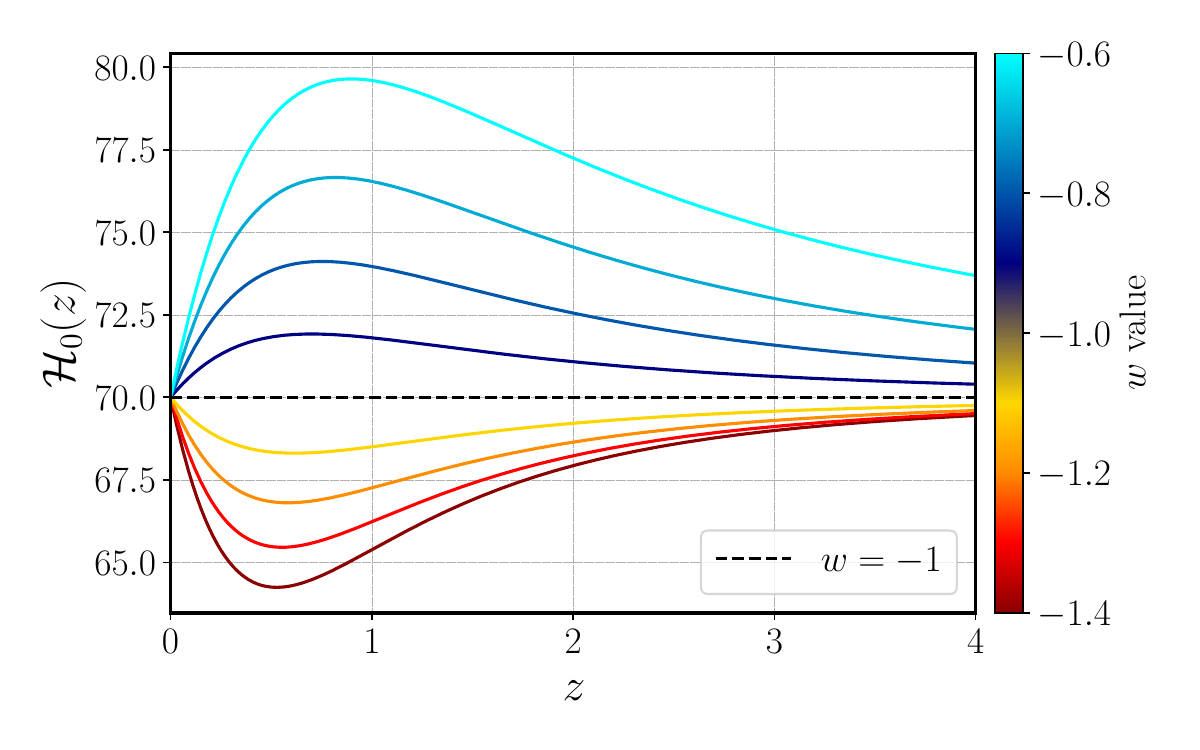}
\includegraphics[width=0.54\textwidth]{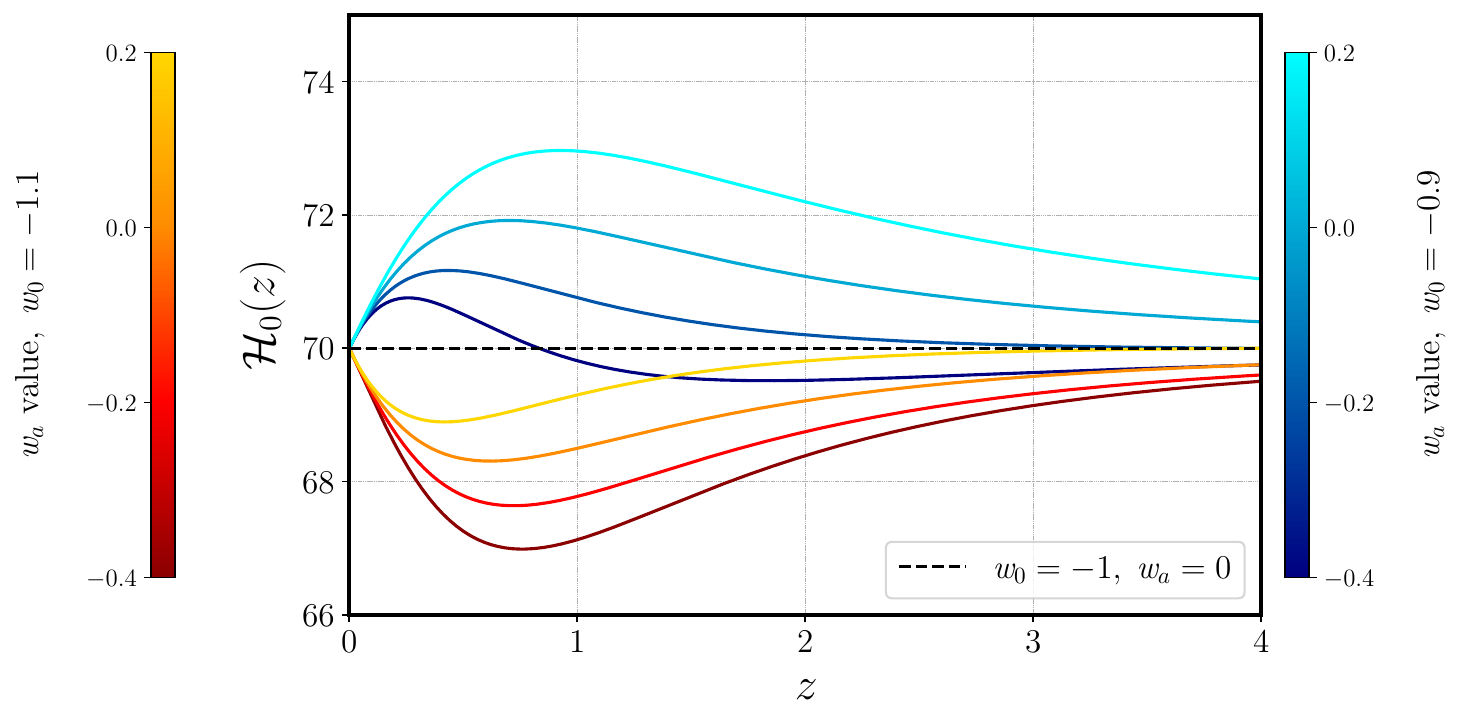}
\caption{Effective running Hubble constant $\mathcal{H}_0(z)$ for two dynamical dark--energy parameterizations. The left panel shows the single--parameter $w\mathrm{CDM}$ case, where blue curves correspond to quintessence-like dark energy ($w>-1$) that makes $\mathcal{H}_0(z)$ initially rise, while red curves correspond to phantom-like dark energy ($w<-1$) that makes it initially fall. The right panel displays $\mathcal{H}_0(z)$ for the $w_0w_a\mathrm{CDM}$ model, where the blue color bar on the right denotes the different values of $w_a$ from $-0.4$ to $0.2$ with present-day quintessence $w_0 = -0.9$, whereas the red color bar in the left indicates the today phantom nature for DE ($w_0=-1.1$) and $w_a$ values from $-0.4$ to $0.2$.}
  \label{fig:H0z_two}
\end{figure}

Although we provide the full analytical forms above, they are algebraically involved. Therefore, we present graphical representations of $\mathcal{H}_0(z)$ for both models in Figure~\ref{fig:H0z_two}, to clearly visualize the redshift dependence and compare different dark energy behaviors.
In the left panel we show the $w$CDM case, detailing the three fundamental scenarios: cosmological constant ($w=-1$), quintessence ($w>-1$), and phantom ($w<-1$). \\In the right side, we examine for the $w_0w_a\mathrm{CDM}$ parametrization four distinct scenarios that result from different combinations of $w_0$ values above or below $-1$ and positive or negative values of $w_a$. In this context, the scenarios are characterized according to whether $w(z)$ is greater than $-1$ (quintessence), equal to $-1$ (cosmological constant), or less than $-1$ (phantom). \\
Based on this figure, we can make the following observations:
\begin{itemize}
\item The sign of the first derivative near $z=0$ clearly distinguishes whether dark energy behaves as quintessence or phantom today. Equivalently, whether $\mathcal{H}_0(z)$ increases or decreases around zero determines the current nature of dark energy. This is valid for both the $w$CDM and $w_0w_a\mathrm{CDM}$ parameterizations;
\item In the $w_0w_a$ parameterization, for cases with $w_0 > -1$, a negative $w_a$ controls the extent to which $\mathcal{H}_0(z)$ decreases. Conversely, a positive $w_a$ induces a peak in the first derivative and results in a less pronounced decreasing trend of $\mathcal{H}_0(z)$. In particular, we stress that, at fixed value of $w_0$, the redshift at which the transition takes place is sensitive to the corresponding value of $w_a$.
\end{itemize}

\begin{figure}[htbp]
  \centering
\includegraphics[width=0.8\linewidth]{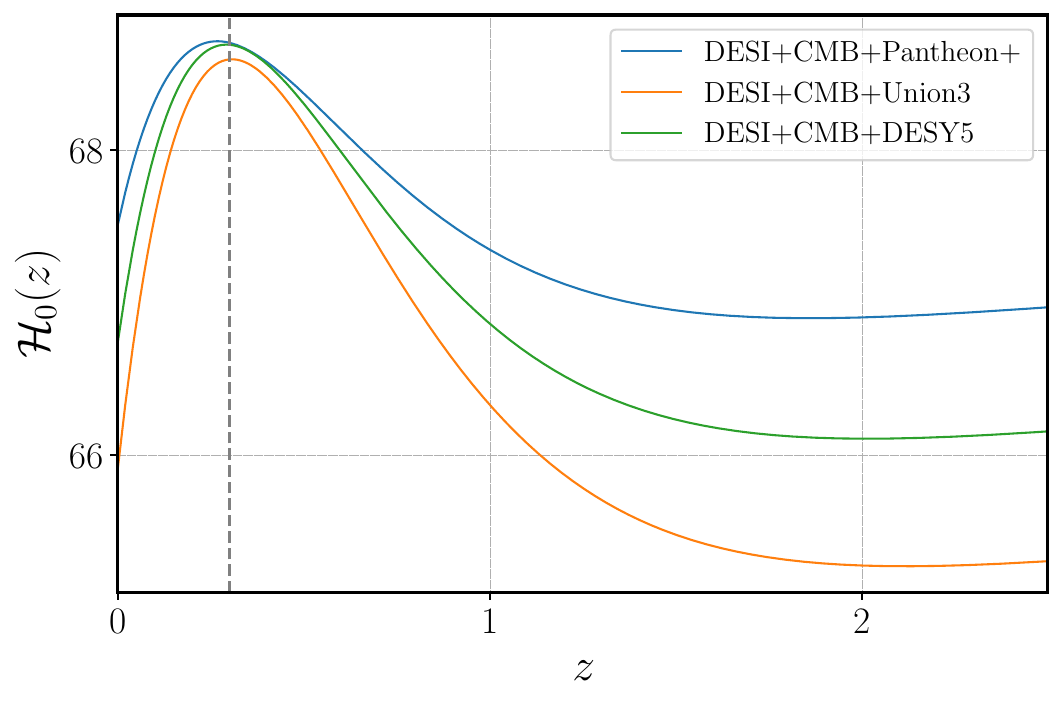}
  \caption{Effective running Hubble constant $\mathcal{H}_0(z)$ for the $w_0w_a\mathrm{CDM}$ model using the DESI best-fit parameters results for the corresponding DESI+CMB+SNe in the three different compilation Pantheon+, Union3 and DESy5.}
  \label{fig:H0z_DESI}
\end{figure}

Notably, recent findings from the DESI collaboration \citep{desi, desi2} have demonstrated that evolving dark energy models are preferred over the cosmological constant. As a modified $\Lambda$CDM model, the DESI Collaboration result naturally features an effective running Hubble constant profile consistent with the scenarios discussed earlier. In Figure~\ref{fig:H0z_DESI}, the three scenarios derived from parameters fitted by DESI (listed in Table V of \citet{desi2}) exhibit a consistent peak around $z\sim0.3$, precisely where dark energy transitions into phantom behavior.

To further clarify the diagnostic power of $\mathcal{H}_0(z)$ in the low-redshift regime, we analyze its analytical behavior at the present time. Based on the expressions derived above for the $\mathcal{H}'_0(z)$ in the $w$CDM and $w_0w_a\mathrm{CDM}$ scenarios, i.e. eqs. \eqref{eq:H0prime_theo_wCDM}, \eqref{eq:H0prime_theo_w0waCDM}, we evaluate them at $z=0$ and obtain, for both models:

\begin{equation}
\mathcal{H}'_{0}(z=0)=\frac{3}{2}H_0(1-\Omega_{m0})(1+w_0),
\label{eq:H0_der_z0}
\end{equation}
from which it immediately follows \footnote{Assuming a universe containing not only matter, as established by \citet{riess1998}} that the sign of $\mathcal{H}'_{0}(z=0)$ directly depends on $(1+w_0)$.  Hence, the effective running Hubble constant emerges as a robust and innovative probe to investigate the fundamental nature of dark energy.

Is it worth noting that the main contribution to the first derivative of $\mathcal{H}_0(z)$ comes from the dark energy parameters. The value of $H_0$ simply fixes the starting value at $z=0$, and variations in $\Omega_{m0}$ (see ~\ref{sec:appA}) do not affect whether $\mathcal{H}_0(z)$ increases or decreases at low redshift; rather, they influence the redshift at which the peak occurs.

\section{Dark energy models}\label{sec3}

In this analysis, we further demonstrate how the effective running Hubble constant can discriminate among different dark energy scenarios by using the data available from a binned Supernovae data analysis as in \citet{dainotti2021, dainotti2022, dainotti_Master}. We consider three different models of dark energy, starting from two phenomenological frameworks, i.e. the $w$CDM and a simplified version of the $w_0w_a\mathrm{CDM}$ model, and concluding with a new theoretical model for dark energy. Each model incorporates only three free parameters, $\Omega_{m0}$, $H_0$ and one additional parameter, i.e. $w$, characterizing the nature of dark energy, due to inherent constraints in binned analyses (see \citet{dainotti_Master, montani-carlevaro-dainotti2024, montani-carlevaro-dainotti2025}).

\subsection{Phenomenological models: \texorpdfstring{$w$CDM}{wCDM} and reduced CPL}\label{subsec:redCPL_model}

\noindent The first model analyzed is the well-known $w$CDM, where the dark energy equation of state is constant, i.e. $w = P/\rho \equiv \mathrm{const.\,}$ . Assuming a flat Universe and considering the eq. \eqref{eq:H_0_z_2}, the associated effective running Hubble constant is given by:

\begin{equation}
\mathcal{H}_0(z)\equiv H_0 \sqrt{\frac{\Omega_{m0}(1+z)^3 + \left( 1 - \Omega_{m0}\right)(1+z)^{3(1+w)}}{\Omega_{m0}(1+z)^3 + \left( 1 - \Omega_{m0}\right)}}\,.
\label{eq:H0z_wCDM}
\end{equation}

The second model is a reduced version of the CPL parametrization, hereafter denoted as CPL', where we impose the condition $w_0 = w_a=w$ to reduce the number of parameters to one. This choice is motivated by one of the DESI-DR2 results (see Table V in \citet{desi2}), where $w_0 \sim w_a$. We emphasize that this simplified case preserves an analytical structure of the dark energy equation of state that remains close to the standard $w_0w_a\mathrm{CDM}$ parametrization. In fact, we chose $w_0=w_a$ because this is the only case in which, given the equality, we can show the phantom vs quintessence crossing and we can test if it is favored by the binned SNe data.

To this end, we recall that the Hubble function for the flat $w_0w_a\mathrm{CDM}$ model is expressed as:

\begin{equation}
H(z)=H_0 \sqrt{\Omega_{m0}(1+z)^3 + \left( 1 - \Omega_{m0}\right)(1+z)^{3(1+w_0+w_a)}e^{-3w_a\frac{z}{1+z}}}\,,
\label{eq:Hz_w0waCDM}
\end{equation}
by setting $w_0 = w_a \equiv w$, we obtain the CPL' version of the Hubble parameter:

\begin{equation}
H(z)=H_0 \sqrt{\Omega_{m0}(1+z)^3 + \left( 1 - \Omega_{m0}\right)(1+z)^{3(1+2w)}e^{-3w\frac{z}{1+z}}}\,.
\label{eq:Hz_w0=waCDM}
\end{equation}
Substituting this expression in eq. \ref{eq:H_0_z_2}, the effective running Hubble constant for the CPL' model becomes:

\begin{equation}
\mathcal{H}_0(z)\equiv H_0 \sqrt{\frac{\Omega_{m0}(1+z)^3 + \left( 1 - \Omega_{m0}\right)(1+z)^{3(1+2w)}e^{-3w\frac{z}{1+z}}}{\Omega_{m0}(1+z)^3 + \left( 1 - \Omega_{m0}\right)}}\,.
\label{eq:H0z_CPL'}
\end{equation}

Notably, the effective running Hubble constant for both the $w$CDM and CPL' models can exhibit a non-monotonic behavior: increasing and then decreasing with redshift, or \textit{vice versa}. This allows in principle for transitions between different dark energy regimes, such as from quintessence-like to phantom-like behavior.

\subsection{Dynamical dark energy model}\label{subsec:dde_model}
\noindent After discussing the phenomenological approaches, we now introduce a novel theoretical model that aims to provide a deeper explanation for the nature of dark energy. \\
In particular, we formulate a theoretical scenario for the late Universe evolution, in which dark energy has an evolutionary character, being created by the cosmological gravitational field. For a detailed discussion see T. Schiavone et al., "Revisiting the Matter-Creation Process: Constraints from Late-Time Acceleration and the Hubble Tension", in preparation.

We consider a flat isotropic Universe \citep{efstathiou_planck} (for a different geometrical setup see \citet{divalentino_planck}, i.e. we deal with a line element of the form

\begin{equation}
	ds^2 = dt^2 - a^2(t)\left( dx^2 + dy^2 + dz^2\right)
	\, ,
	\label{dem1}
\end{equation}
where $t$ denotes the synchronous time (we consider $c=1$ units), $x,y,z$ are Cartesian coordinates and $a(t)$ is the cosmic scale factor, regulating the Universe expansion process. \\
We describe the late Universe dynamics as characterized by the presence of the (dark and baryonic) matter component $\rho_m$, which retains the standard morphology and the dark energy density $\rho_ {de}$, which is associated to a constant equation of state parameter $w_{de}$ and which is created by the cosmological gravitational field via a negative pressure term $-p_c$, taking the form \citep{matcre_calvaoLima, matcre_montani2001, elizalde_odintsov}:

\begin{equation}
	3Hp_c\equiv -\Gamma 
	\left( 1 + w_{de}\right) \rho_{de}
	\, , 
	\label{dem2}
\end{equation}
where $\Gamma$ is a phenomenological constant matter creation rate.
Hence, the dark energy evolution is governed by the following modified continuity equation:

\begin{equation}
	\dot{\rho}_{de}=-3H\left[ 
	(1+w_{de})\rho_{de} + p_c\right]
	\, , 
	\label{dem3}
\end{equation}
where the dot denotes time differentiation  with respect to $t$. For similar works see \citet{matcre_montanimary, fazzari_deleo}.

Passing to the redshift time variable $z(t) = a_0/a-1$ ($a_0$ being the present-day value of the scale factor), 
so that $\dot{(...)}=-(1+z) H(...)^{\prime}$, the prime referring to $z$-differentiation, the dynamical scheme of the proposed model is summarized by the Friedmann equation 

\begin{equation}
	H^2(z) = \frac{\chi}{3}\left( 
	\rho_m(z) + \rho_{de}(z)\right)
	\, 
	\label{dem4}
\end{equation}
($\chi$ being the Einstein constant) and the two continuity equations for the matter

\begin{equation}
	\rho_m^{\prime}=\frac{3}{1+z}\rho_m
	\, 
	\label{dem5} 
\end{equation}
and for the dark energy

\begin{equation}
	\rho_{de}^{\prime} = 
	\frac{3}{1+z}(1+w_{de}) 
	\left( 1 - \frac{\Gamma}
	{3H}\right) \rho_{de}
	\,  , 
	\label{dem6}
\end{equation}
respectively.
These two continuity equations admit the explicit solutions:

\begin{equation}
	\rho_m(z)=\rho_m^0(1+z)^3 
	\quad ,\, 
	\rho_{de}(z) = \rho^0_{de}
	(1+z)^{3(1+w)}\exp 
	\left\{ -(1+w) \gamma_0 \int_0^z\frac{dy}{(1+y)E(y)} \right\}
	\, , 
	\label{dem7}
\end{equation}
where $E\equiv H(z)/H(z=0)$, $w \equiv w_{de}$ and 
$\gamma_0\equiv \Gamma /H(z=0)$.  Here, the label $0$ indicates the value, taken by the two energy densities today, i.e. in $z=0$. 

Introducing the following standard definitions and normalization condition:

\begin{equation}
	H_0\equiv H(z=0) \, ,\, 
	\Omega_{m0}\equiv \frac{\chi\rho_{dm}^0}{3H_0^2}
 ,\, 
	\Omega_{de}^0 \equiv 
	\frac{\chi\rho_{de}^0}{3H_0^2} \equiv
	1 - \Omega_{m0}
	\, , 
	\label{dem8}
\end{equation}
then, Eq.\eqref{dem4} can be restated as: 

\begin{equation}
	E^2(z) = \Omega_{m0}(1+z)^3 + \left( 1 - \Omega_{m0}\right)(1+z)^{3(1+w)}\exp \left\{ -(1+w) \gamma_0 \int_0^z\frac{dy}{(1+y)E(y)}\right\}
	\, .
	\label{dem9}
\end{equation}
Thus, we see that, our model, corresponds to an evolutionary dark energy scenario, according to the analysis of the DESI Collaboration \citep{desi, desi2}, see also \citet{colgain_Dainotti, Giare_Robust_DDE}. The expansion rate above contains three free parameters, i.e. $\Omega_{m0}$, $w$ and $\gamma_0$, to which we have to add the value of $H_0$, when the Hubble parameter is reconstructed. 

In order to ensure that the proposed model behaves well near $z\simeq0$ and, in particular, that it fulfills the luminosity-redshift relation \citep{riess2016, efstathiou2021}, we analyze the cosmographic parameters \citep{weinberg-grav-cosm}. 
It is easy to check that the deceleration parameter $q_0$ can be stated as follows: 

\begin{equation}
	q_0\equiv - 1 + \frac{1}{2}(E^2)^\prime_{\mid_{z=0}}
	\,  .
	\label{dem10}
\end{equation}
Hence, according to Eq.\eqref{dem9}, we get:

\begin{equation}
q_0 = - 1 + \frac{3}{2}\Omega_{m0} + \frac{3(1-\Omega_{m0})}{2}(1+w) \left( 1 - \frac{\gamma_0}{3}\right) = q_0^{\Lambda CDM} + \frac{3(1-\Omega_{m0})}{2}(1+w)\left( 1 - \frac{\gamma_0}{3}\right) \, , 
\label{dem11}
\end{equation}

\noindent where $q_0^{\Lambda CDM}$ denotes the standard $\Lambda$CDM value of the deceleration parameter. Thus, in order to remain close to this value, in what follows, we require the condition $\gamma _0=3$. \\
Therefore, in this limit, Eq.\eqref{dem9} reduces to:

\begin{equation}
    E^2(z)=\Omega_{m0}(1+z)^3 + \left( 1 - \Omega_{m0}\right)(1+z)^{3(1+w)}\exp \left\{ -3(1+w) \int_0^z\frac{dy}{(1+y)E(y)}\right\}\,.
    \label{dem14}
\end{equation}
and given this modified Hubble parameter, we can derive the corresponding theoretical running Hubble constant function by substituting this expression into Eq.\eqref{eq:H_0_z_2}, yielding:

\begin{equation}
        \mathcal{H}_0(z)\equiv H_0 \sqrt{\frac{\Omega_{m0}(1+z)^3 + \left( 1 - \Omega_{m0}\right)(1+z)^{3(1+w)}\exp \left\{ -3 (1+w) \int_0^z\frac{dy}{(1+y)E(y)}\right\}}{\Omega_{m0}(1+z)^3 + \left( 1 - \Omega_{m0}\right)}} \,.
        \label{eq:H0z_DDE}
\end{equation}

\section{Statistical analysis}\label{sec4}
\noindent We analyze how the binned SNe Ia samples can constrain the theoretical effective running Hubble constant for the proposed models, discerning the nature of dark energy as either quintessence or phantom. First, we present the analysis conducted using both the full and binned SNe Ia datasets. Subsequently, for completeness, we show the results of the background analysis, allowing all cosmological parameters to vary freely within predefined flat priors. 

\subsection{Datasets}
\noindent The binned data used in this article are derived by analyzing SNe Ia observations in redshift intervals. Specifically, we use two distinct SNe Ia samples for the binned analysis. Here, we briefly recall how these binned samples are constructed. In each redshift bin, the theoretical distance modulus, $\mu_{th}=5\log_{10}d_L(z; H_0, \Omega_{m0})+25$, is computed assuming a baseline $\Lambda$CDM cosmological model and compared with the observed distance modulus $\mu_{obs}=m_B - M$. This procedure yields one effective value of $H_0$ and $\Omega_{m0}$ \textit{per} bin. 
To extract these values, a Markov Chain Monte Carlo (MCMC) analysis is carried out independently for each bin, allowing $H_0$ and $\Omega_{m0}$ to vary within dataset-specific prior ranges (see Table~\ref{tab:prior_compl}). It is worth noting that each extracted $H_0$ value in a given redshift bin is independent from the others and also from the previous at lower redshifts, in this sense each bin is treated as statistically independent, with no correlations between bins assumed. The calibration is performed using $H_0 = 70$ km/s/Mpc since this choice does not affect the qualitative features of the effective running Hubble constant, which is the focus of our study. 
Therefore, the two binned samples are:

\begin{itemize}

    \item \textit{Master binned SNe Ia sample} - the new binned SNe Ia sample \citep{dainotti_Master} that merges 3714 SNe Ia from Dark Energy Survey (DES) \citep{DESy5}, PantheonPlus \citep{Scolnic_2022, Brout_2022}, Pantheon \citep{scolnic2018} and Joint Lightcurve Analysis (JLA) \citep{JLA} without duplicates. Specifically:
    \begin{itemize}
    \item DES contributes all 1829 available SNe Ia;
    \item PantheonPlus provides 1208 SNe Ia after removing internal duplicates (158) and overlaps with DES (335);
    \item Pantheon adds 181 SNe Ia after removing 867 duplicates with Pantheon+ and DES;
    \item JLA includes 496 SNe Ia after removing 244 duplicates shared with other catalogs.
    \end{itemize}
    The final dataset used in this work is arranged into 20 equally populated bins, spanning redshifts from $0.0012$ to $2.3$. 
    As reported in Table \ref{tab:binned_intervals}, we note that the mean redshift of the first bin is $0.0091$, while the mean redshift of the last bin is $1.54$. 
    The creation of this binned sample is based on the procedure described in \citet{dainotti_Master}, where a flat prior on $H_0$ within the range $[60, 80]$ was assumed, together with a Gaussian prior at the $5\sigma$ level for $\Omega_{m0}=0.322 \pm 0.025$, corresponding to the fiducial mean value derived from fitting the complete sample within the $\Lambda$CDM model. The PL fit provides best-fit parameters $\alpha=0.010$ and $H_0=69.869$ km/s/Mpc. 
    This dataset is referred to as `Master bin.'.

    \item \textit{Pantheon binned SNe Ia sample} - that consists of 20 equally populated bins made up of $\sim 52$ SNe Ia \citep{dainotti_Master}, resulting in a range of redshift of $ 0.016 < z < 1.56$, corresponding to the mean values of the redshift in each bin. Details about the redshift intervals are shown in Table \ref{tab:binned_intervals}. The construction of this binned sample follows the analysis of \citet{dainotti_Master}, where a flat prior for $H_0$ in the interval $[60, 80]$ and a Gaussian prior at 1$\sigma$ for $\Omega_{m0}=0.298 \pm 0.022$ were adopted. Since the Pantheon has a reduced sample size, being $28\%$ of the Master, this allows us to adopt more informative priors at 1 $\sigma$. 
    Following ~\cite{dainotti2021}, the bin-dependent $H_0$ values are fitted using a phenomenological power-law (PL) parametrization, $\mathcal{H}_0(z)=H_0/(1+z)^\alpha$, yielding best-fit parameters $\alpha=0.0169$ and $H_0=70.198$ km/s/Mpc. We denote this sample as `Pantheon bin'.
\end{itemize}

\noindent The choice to use only 20 bins for our purpose (versus the original 3,12 and 20 bins cases presented in \citet{dainotti_Master}) optimizes statistical precision while minimizing binning-induced uncertainties. 

As additional background data, we include:
\begin{itemize}
    \item \textit{Type Ia Supernovae}: Distance moduli from two major compilations:
    \begin{itemize}
    \item[-] Pantheon sample \citep{scolnic2018}, containing 1048 SNe Ia covering $0.01 < z < 2.3$. We denote this dataset as `Pantheon'.
    \item[-] Master sample \citep{dainotti_Master} combining DES, PantheonPlus, Pantheon and JLA SNe Ia without duplicates. We label this dataset as `Master'.
    \end{itemize}
    
\item \textit{DESI-DR2 BAO}: Measurements of Baryon Acoustic Oscillations (BAO) from DESI's second-year release \citep{desi2}, spanning $0.1 < z < 4.2$, calibrated to Planck's sound horizon $r_d = (147.09\pm0.26)$ Mpc as reported in Table 2 of Ref. \citet{Planck2018}. We refer to this dataset as `DESI'.

\item \textit{Cosmic Chronometers (CC)}: Expansion rate measurements $H(z)$ from differential galaxy ages \citep{Jimenez_247, CC_borghi}. Our analysis focuses on a subset of 15 points \citep{Moresco_2012, moresco_2015, Moresco_2016} in the redshift interval $0.1791 < z < 1.965$. This dataset is referred to as `CC'.
\end{itemize}

To perform these analyses we used the publicly available sampler \texttt{Cobaya} \citep{cobaya} to implement MCMC as sampler method to compare theoretical functions against data. We adopted the Gelman-Rubin criterion \citep{gelman_rubin} to assess convergence of our chains, i.e. $R-1 < 0.01$. The statistical results and plots are obtained with \texttt{getdist} software \citep{getdist}. Specifically, we used a preliminary version of a code that will soon be publicly released (Giarè, Fazzari, in prep.). 
In particular, to fit the theoretical $\mathcal{H}_0(z)$ function to the corresponding binned data samples, the likelihood is constructed in order to minimize the total chi-squared:
\begin{equation}
    \chi^2_{tot}=\chi^2_{H_{0_{bin}}}+\chi^2_{\Omega_{m0_{bin}}} \,,
\end{equation}
where $X_{\text{bin}}$ denotes the single data value in each bin, i.e., the binned values of $H_0$ and $\Omega_{m0}$. \\
Regarding prior ranges, for the background analysis we adopted broad flat priors for all the free parameters. To ensure consistency when comparing the binned and full SNe Ia samples, we applied a flat prior on the single free parameter describing the dark energy equation of state $w$, while for $\Omega_{m0}$ and $H_0$ we used the same prior ranges as in the binned data analysis of \citet{dainotti_Master}, applying them consistently to both the binned and full SNe Ia analyses. All the prior ranges adopted are summarized in Table~\ref{tab:prior_compl}.

We have also tested our models using the Levenberg–Marquardt algorithm, and found that the resulting best-fit values are consistent with the posterior maxima obtained from the MCMC analysis. 

For model selection, we compute the differences in the logarithm of the Bayes factor, defined as $\ln B_{i,j} = \ln \mathcal{Z}_j - \ln \mathcal{Z}_i$ \citep{Bayes_trotta}, where $\mathcal{Z}_i$ and $\mathcal{Z}_j$ denote the Bayesian evidence for models $i$ and $j$, respectively. The Bayesian evidence quantifies how well a model fits the data while penalizing excessive complexity or a larger number of free parameters. This calculation is performed using the Cobaya wrapper available at \citet{William_git}, transforming the Gaussian priors into uniform distributions applying the same approach followed in \citet{Favale_uniform}. To interpret the significance of this factor, we adopted Jeffrey’s scale \citep{jeffreys_scale}, which categorizes the evidence against the model as inconclusive if $0 < |\ln B_{i,j}| < 1.0$, weak if $1.0 < |\ln B_{i,j}|< 2.5$, moderate if $2.5 < |\ln B_{i,j}| < 5.0$, and strong if $|\ln B_{i,j}| > 5.0$.

\begin{table}[ht!]
\centering
\begin{tabular}{llllll}
\toprule
\multicolumn{4}{c}{\shortstack[l]{\centering{Prior for different datasets}}} \\
\midrule\midrule
Parameter & Background & Pantheon  & Master \\
\midrule
$\Omega_{m0}$ & $\mathcal{U}[0.01,0.99]$ & $\mathcal{N}[0.298,0.022]$ &  $\mathcal{N}[0.322,0.025]$\\
$H_0 \,\mathrm{(km/s/Mpc)}$ & $\mathcal{U}[20,100]$   & $\mathcal{U}[60,80]$ &  $\mathcal{U}[60,80]$ \\
$w$     & $\mathcal{U}[-7,3]$     & $\mathcal{U}[-7,3]$ & $\mathcal{U}[-7,3]$ \\
\bottomrule
\end{tabular}
\caption{Prior ranges imposed for the free parameters of the analyzed models. Same priors of Pantheon and Master are used also for binned Pantheon and binned Master analyses.}
\label{tab:prior_compl}
\end{table}

\begin{table}[ht!]
  \centering
  \small
  \begin{subtable}[t]{0.43\textwidth}
    \centering
    \begin{tabular}{cc}
      \toprule
      $\bar{z}$ & Interval\\ 
      \midrule
      0.0091 & [0.0012,\,0.0169] \\
      0.0218 & [0.0169,\,0.0267] \\
      0.0321 & [0.0267,\,0.0375] \\
      0.0561 & [0.0376,\,0.0746] \\
      0.1066 & [0.0749,\,0.1382] \\
      0.1594 & [0.1383,\,0.1806] \\
      0.1978 & [0.1806,\,0.2149] \\
      0.2319 & [0.2151,\,0.2486] \\
      0.2664 & [0.2487,\,0.2841] \\
      0.3023 & [0.2842,\,0.3204] \\
      0.3366 & [0.3204,\,0.3529] \\
      0.3771 & [0.3530,\,0.4013] \\
      0.4268 & [0.4019,\,0.4518] \\
      0.4788 & [0.4519,\,0.5056] \\
      0.5292 & [0.5057,\,0.5527] \\
      0.5769 & [0.5528,\,0.6009] \\
      0.6269 & [0.6015,\,0.6520] \\
      0.6858 & [0.6520,\,0.7191] \\
      0.7711 & [0.7195,\,0.8227] \\
      1.5430 & [0.8239,\,2.2610] \\
      \bottomrule
    \end{tabular}
    \caption{Master binned dataset}
    \label{tab:master_interval}
  \end{subtable}
  \hfill
  \begin{subtable}[t]{0.43\textwidth}
    \centering
    \begin{tabular}{cc}
      \toprule
      $\bar{z}$ & Interval \\ 
      \midrule
      0.0155 & [0.0101,\,0.0209] \\
      0.0265 & [0.0212,\,0.0318] \\
      0.0407 & [0.0319,\,0.0495] \\
      0.0718 & [0.0498,\,0.0939] \\
      0.1116 & [0.0940,\,0.1292] \\
      0.1431 & [0.1294,\,0.1568] \\
      0.1687 & [0.1568,\,0.1806] \\
      0.1917 & [0.1806,\,0.2028] \\
      0.2130 & [0.2029,\,0.2231] \\
      0.2351 & [0.2237,\,0.2465] \\
      0.2582 & [0.2481,\,0.2683] \\
      0.2820 & [0.2685,\,0.2956] \\
      0.3110 & [0.2965,\,0.3255] \\
      0.3430 & [0.3260,\,0.3600] \\
      0.3894 & [0.3602,\,0.4187] \\
      0.4583 & [0.4193,\,0.4974] \\
      0.5447 & [0.5007,\,0.5887] \\
      0.6602 & [0.5903,\,0.7301] \\
      0.7938 & [0.7336,\,0.8540] \\
      1.5593 & [0.8585,\,2.2600] \\
      \bottomrule
    \end{tabular}
    \caption{Pantheon binned dataset}
    \label{tab:Pantheon_interval}
  \end{subtable}
  \caption{Mean redshift $\bar z$ and redshift binning intervals for the Master and Pantheon binned datasets.}
  \label{tab:binned_intervals}
\end{table}

\section{Results}\label{sec5}
\noindent In this section, we present the cosmological constraints derived from our analysis of the DE models: $w$CDM, CPL' (with $w_0=w_a$) and DDE, using both binned and full data samples from Pantheon and Master datasets, alongside with a background analysis from additional datasets (DESI+CC).

\subsection{Constraints from binned and full SNe Ia data samples}\label{sub:5.1}
\noindent In Table~\ref{tab:binned_full}, we summarize the inferred cosmological parameters from the MCMC analysis for each DE model using both binned and full SNe Ia datasets. The parameters considered include the Hubble constant ($H_0$), matter density ($\Omega_{m0}$), and DE equation of state (EoS) parameter ($w$). 
It is worth noting that, following \citet{dainotti_Master}, the Pantheon and Master binned samples were divided into redshift bins and fitted to $\Lambda$CDM to obtain $H_0$ values, interpreted as observational estimates of the effective running Hubble constant $\mathcal{H}_0(z)$. In our work, these estimates are compared with the theoretical $\mathcal{H}_0(z)$ functions of the three DE models, i.e. Eqs.~\eqref{eq:H0z_wCDM}, \eqref{eq:H0z_CPL'}, and \eqref{eq:H0z_DDE}, allowing us to perform a single global MCMC fit across all bins with $H_0, \, \Omega_{m0},\, w$, as free parameters of the models.

\begin{table}[ht!]
\centering
\renewcommand{\arraystretch}{1.6}  
\resizebox{\textwidth}{!}{
\begin{tabular}{
>{\centering\arraybackslash}p{1.5cm} 
>{\centering\arraybackslash}p{1.5cm} 
c c c c c c}
   \hline
     & & \multicolumn{4}{c}{Datasets} \\
    \cline{3-6}
     &  & \textbf{Master bin.} & \textbf{Pantheon bin.} & \textbf{Master} & \textbf{Pantheon} \\
    \hline
    \hline
    \multirow{4}{*}{\parbox[c]{1.5cm}{\centering\rotatebox{90}{$w$CDM}}}
    & \parbox[c]{1.5cm}{\centering{$H_0$}} 
    & $  69.865\pm 0.083 $ & $  70.53\pm 0.27 $ & $  69.11\pm 0.40 $ & $  70.08\pm 0.45 $  \\
    & \parbox[c]{1.5cm}{\centering{$\Omega_{m0}$}} 
    & $  0.3245\pm 0.0053 $ & $  0.2984\pm 0.0047 $ & $  0.318\pm 0.024 $ & $  0.306^{+0.040}_{-0.036} $ \\
    & \parbox[c]{1.5cm}{\centering{$w$}} 
    & $  -1.012\pm 0.014 $ & $  -1.052\pm 0.025 $ & $  -0.904\pm 0.067 $ & $  -1.04^{+0.13}_{-0.11} $   \\
    \cline{2-6}
    & \parbox[c]{1.5cm}{\centering{$\epsilon_w$ (\%)}} 
    & $1.38\%$ & $2.38\%$ & $7.41\%$ & $11.54\%$ \\
    \hline
    \hline
    \multirow{4}{*}{\parbox[c]{1.5cm}{\centering\rotatebox{90}{CPL$'$}}}
    & \parbox[c]{1.5cm}{\centering{$H_0$}} 
    & $ 69.609\pm 0.078 $ & $  69.85\pm 0.27 $ & $  68.92^{+0.35}_{-0.39} $ & $  69.74\pm 0.42$ \\
    & \parbox[c]{1.5cm}{\centering{$\Omega_{m0}$}} 
    &   $ 0.3257\pm 0.0053 $ & $  0.2995\pm 0.0048 $ & $  0.328\pm 0.023 $ & $  0.305\pm 0.022 $ \\
    & \parbox[c]{1.5cm}{\centering{$w$}} 
    & $  -0.884\pm 0.012 $ & $  -0.884\pm 0.023 $ & $  -0.808^{+0.065}_{-0.057} $ & $  -0.889^{+0.081}_{-0.065} $   \\
    \cline{2-6}
    & \parbox[c]{1.5cm}{\centering{$\epsilon_w$ (\%)}} 
    & $1.36\%$ & $2.60\%$ & $7.55\%$ & $8.21\%$ \\
    \hline
    \hline
     \multirow{4}{*}{\parbox[c]{1.5cm}{\centering\rotatebox{90}{DDE}}} 
    & \parbox[c]{1.5cm}{\centering{$H_0$}} 
    & $  69.836\pm 0.062 $ & $  70.16\pm 0.17 $ & $  69.50\pm 0.30 $ & $  70.01\pm 0.31 $ \\
    & \parbox[c]{1.5cm}{\centering{$\Omega_{m0}$}} 
    & $  0.3243^{+0.0048}_{-0.0055} $ & $  0.2982\pm 0.0048 $ & $  0.343\pm 0.024 $ & $  0.302^{+0.021}_{-0.024} $ \\
    & \parbox[c]{1.5cm}{\centering{$w$}} 
    & $  -1.12\pm 0.14 $ & $  -1.21^{+0.19}_{-0.17} $ & $  -1.06^{+0.83}_{-0.60} $ & $  -1.19^{+0.92}_{-0.63} $ \\
    \cline{2-6}
    & \parbox[c]{1.5cm}{\centering{$\epsilon_w$ (\%)}} 
    & $12.5\%$ & $14.9\%$ & $67.5\%$ & $65.1\%$ \\
    \hline
\end{tabular}
}
\caption{Mean values and associated uncertainties for the inferred parameters from the MCMC analysis for the three DE models: $w$CDM, the reduced CPL' ($w_0=w_a$) and the theoretical DDE, presented for binned SNe data samples and full SNe datasets. Relative errors on DE parameter $w$ ($\epsilon_w$) are also displayed.}
\label{tab:binned_full}
\end{table}

\begin{figure}[ht!]
\centering
\includegraphics[width=0.48\textwidth]{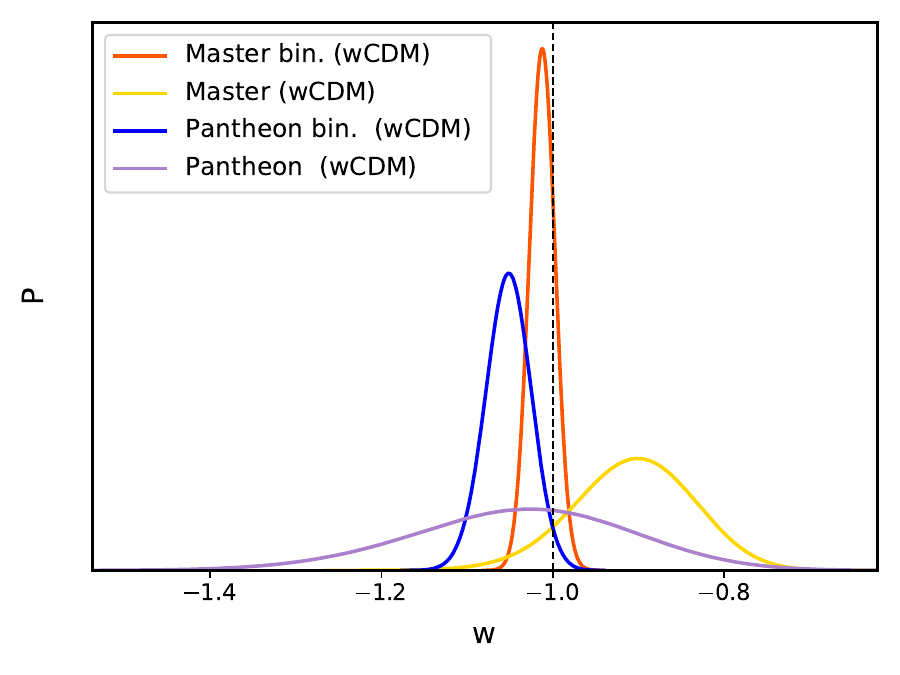}
\includegraphics[width=0.48\textwidth]{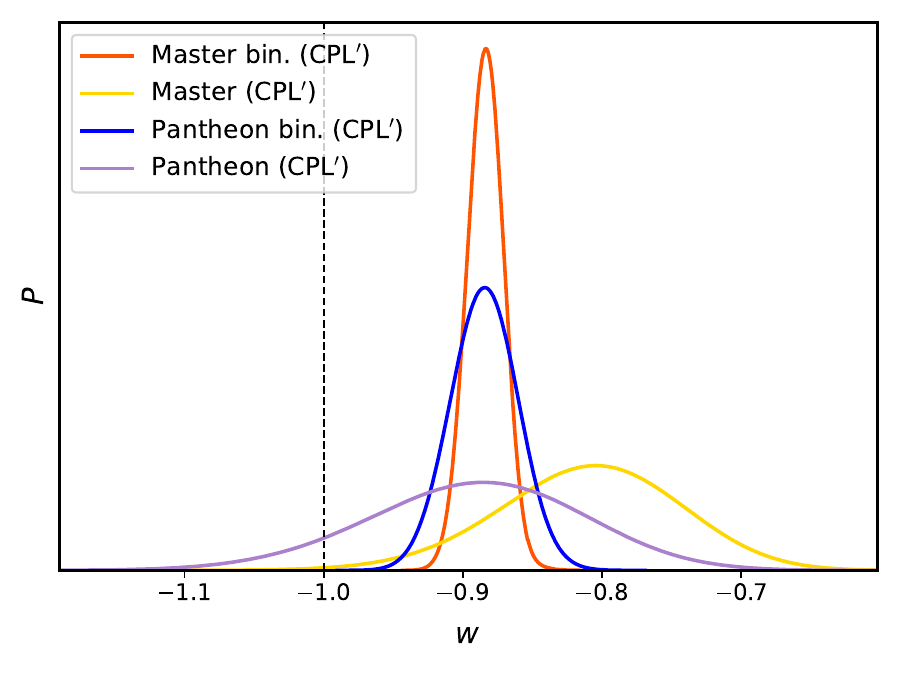}
\includegraphics[width=0.48\textwidth]{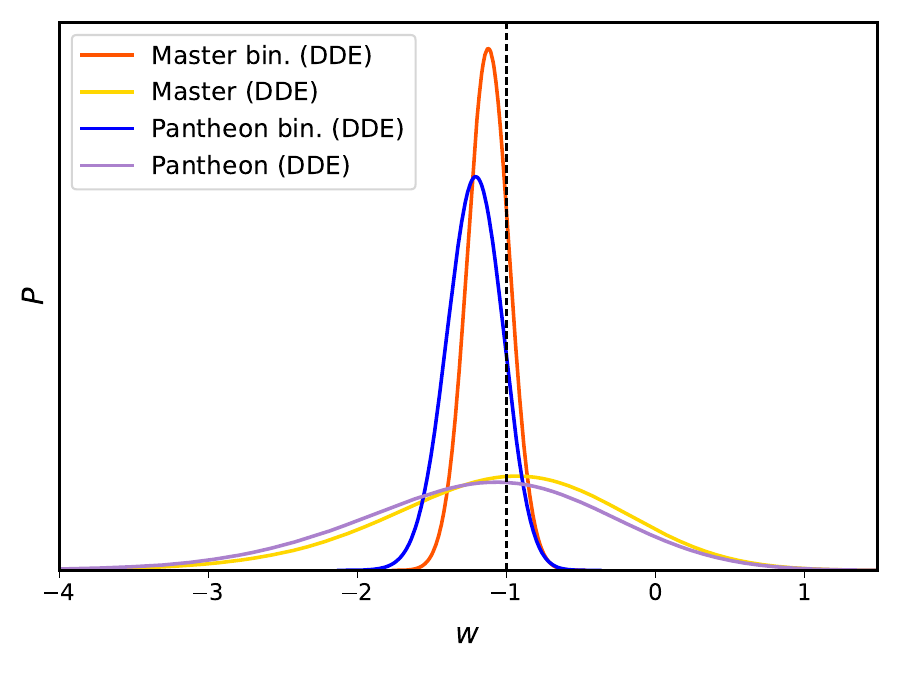}
\caption{One dimensional marginalized probability on DE parameter $w$ obtained using the Master and Pantheon binned data samples and the respective full datasets. The upper-left panel shows the $w$CDM model, the upper-right corresponds to the reduced CPL$^\prime$ case ($w_0 = w_a$), and the lower panel displays the DDE model. The black dashed line represents the cosmological constant case $w = -1$.}
  \label{fig:w_three}
\end{figure}

Figure~\ref{fig:w_three} shows marginalized constraints on $w$ comparing the binned and full SNe Ia samples. \\
We observe that:

\begin{itemize}
        \item Using binned data makes our measurement of $w$ much more precise than using the full Supernovae samples. For instance, in the $w$CDM model we obtain a $1.4 \%$ error on $w$ with the Master binned data (vs. $7.4 \%$ with the full Master sample) and a $2.4 \%$ error with the Pantheon binned data (vs. $11.5 \%$ with the full Pantheon sample). Thus, fitting the theoretical $\mathcal{H}_0(z)$ with binned data  allows us to constrain the nature of dark energy more tightly; 
        \item Furthermore, the Pantheon binned analysis is capable of distinguishing, at least at $1\,\sigma$ level (in the case of $w$CDM greater that $1\,\sigma$), whether the behavior is quintessence-like or phantom-like. In particular, the $w$CDM model indicates a phantom nature of dark energy ($w < -1$), and the DDE model yields a similar result. By contrast, the CPL' model suggests a scenario closer to quintessence ($w > -1$) at low redshifts and a transition to the phantom dark energy at $z_c \sim 0.2$. 
        On the other hand, the Master seems not to be able to constrain between the phantom and quintessence models. One has to note that the Pantheon provides only a contribution of 181 over 3714 SNe Ia in the Master catalog, thus a percentage of only $4.9 \%$. 
        We emphasize that these results are achieved considering an uniform analysis with identical prior ranges and same calibrations used in both cases;
        \item The Master binned dataset provides smaller uncertainties than the Pantheon binned sample, resulting in more precise parameter estimates.
    \end{itemize}

\begin{figure}[ht!]
\centering
\includegraphics[width=0.48\textwidth]{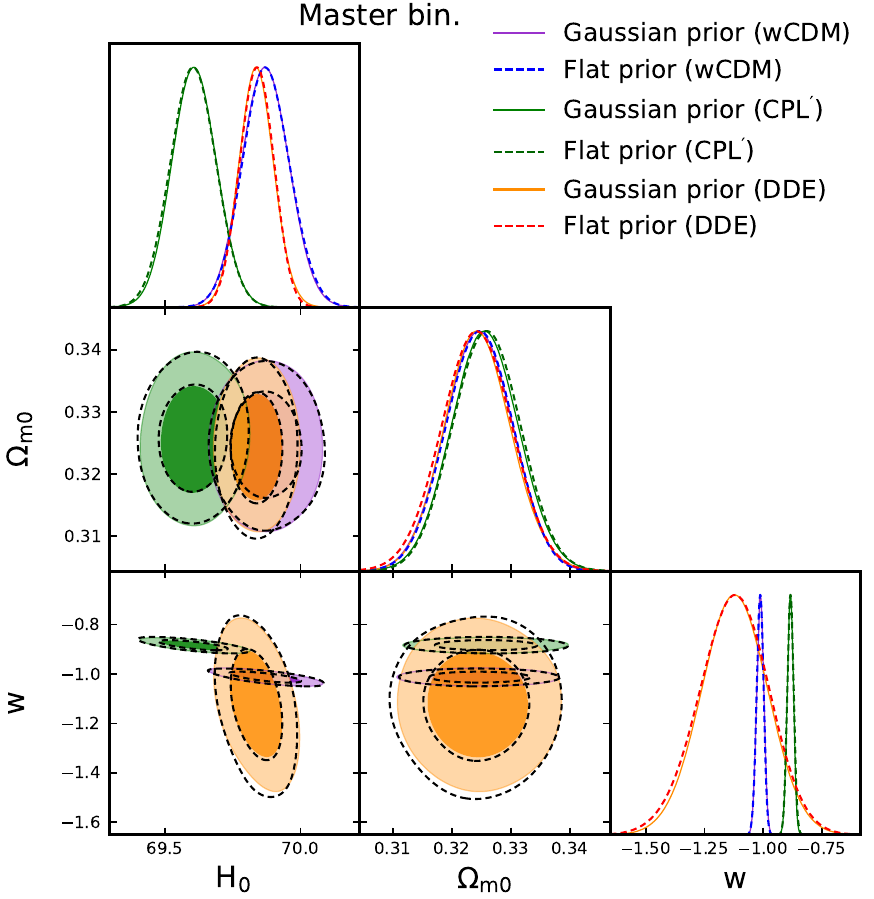}
\includegraphics[width=0.48\textwidth]{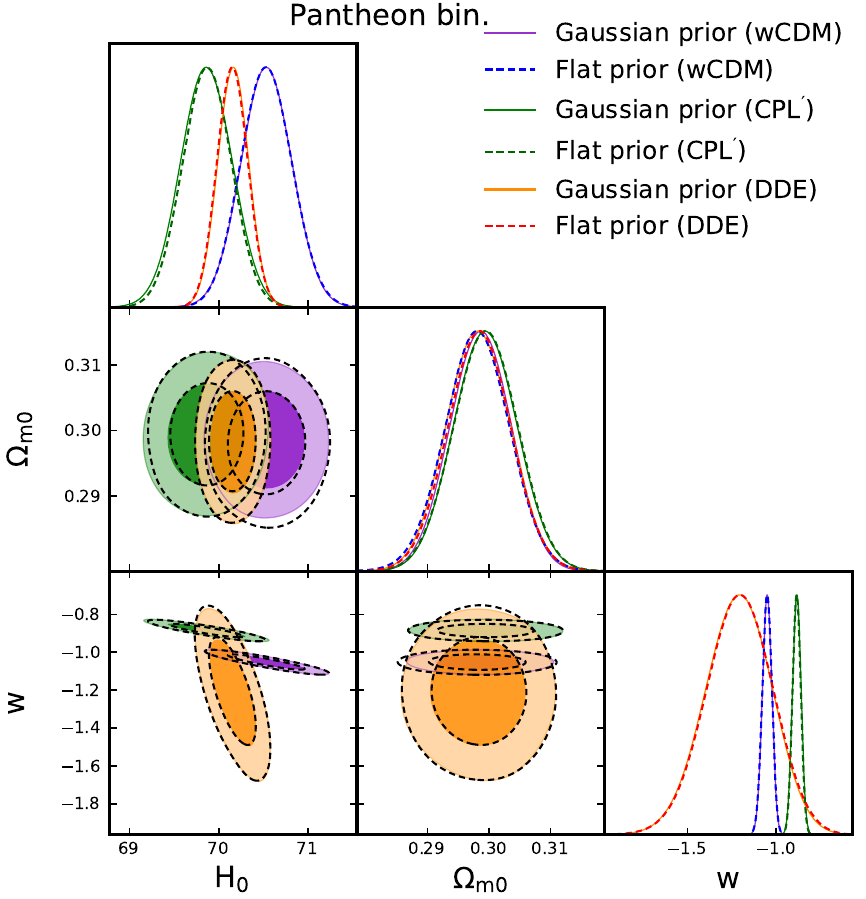}
\caption{One-dimensional posterior distributions and two-dimensional $68\%$ and $95\%$ confidence level contours for the cosmological parameters, derived from the analysis of the Master (left panel) and Pantheon binned (right panel) samples. Results are shown for all three dark energy models ($w$CDM, CPL' and DDE), considering both a Gaussian prior on $\Omega_{m0}$ (see Table~\ref{tab:prior_compl}) and a flat prior throughout $\Omega_{m0}$ $[0.01, 0.99]$, indicating that the parameter constraints from the binned analysis are insensitive to the choice of prior on $\Omega_{m0}$.} 
  \label{fig:priors}
\end{figure}

We stress that we have verified the impact of the prior assumptions on the binned analysis: replacing the Gaussian priors on $\Omega_{m0}$ with flat priors do not significantly affect the inferred values of the three free parameters. The left and right panel of Figure~\ref{fig:priors} show the impact of the priors for the Master and the Pantheon samples, respectively.

 Regarding model selection, in our analysis, we use two reference models as the baseline for comparison. The first is the $\Lambda$CDM, which helps us test whether DE models are favored over the standard cosmological constant scenario. The second is the power-law (PL) model, inspired by previous studies \citep{dainotti2021, dainotti_Master}, where this parametrization was shown to provide a better fit to binned datasets representing the effective running Hubble constant. 
 Looking at the Bayes statistical analysis presented in  Table~\ref{tab:statistics}, we can observe the following:
      \begin{itemize}
          \item Among the dark energy models, $w$CDM and DDE are statistically equivalent and both perform better than the CPL'. Both yield mainly $\ln B_{i,\mathrm{ref}}$ between 2.5 and 5, thus they are moderate disfavored compared to the two ref models i.e. the $\Lambda$CDM and the PL model;
         \item The power-law (PL) model consistently emerges as the most favored across both datasets, with a negative logarithm of the Bayes factor;
         \item The CPL' model is significantly disfavored compared to $\Lambda$CDM, with \\
         $\ln B_{\mathrm{CPL',ref}}>5$, indicating less statistical support for this scenario.
      \end{itemize}

\begin{table}[ht!]
\centering
\begin{tabular}{l rr| rr}
\toprule
 & \multicolumn{2}{c}{\textbf{Master bin.}} & \multicolumn{2}{c}{\textbf{Pantheon bin.}} \\
\cmidrule(lr){2-3}\cmidrule(lr){4-5}
Model 
  & $\ln\mathrm{B}_{i,\Lambda\mathrm{CDM}}$ 
  & $\ln\mathrm{B}_{i,PL}$ 
  & $\ln\mathrm{B}_{i,\Lambda\mathrm{CDM}}$ 
  & $\ln\mathrm{B}_{i,PL}$ \\
\midrule
$w$CDM                 
  &  4.62 &  6.13 
  &  2.45 &  4.53 \\
CPL'      
  & 14.23 & 15.92 
  & 14.72 & 16.75 \\
DDE                    
  &  3.01 &  4.72 
  &  2.76 &  4.61 \\
PL                     
  &  -1.34 & —    
  &  -2.23 & —      \\
\bottomrule
\end{tabular}
\caption{Differences in the logarithmic of Bayesian factor for the three dark-energy models relative to the $\Lambda$CDM and PL reference models, shown separately for the Master and Pantheon binned datasets. Positive values indicate a preference for the reference model, i.e. the $\Lambda$CDM and the phenomenological power-law (PL) model.}
\label{tab:statistics}
\end{table}

\begin{figure}[ht!]
\centering
\includegraphics[width=1.0\linewidth]{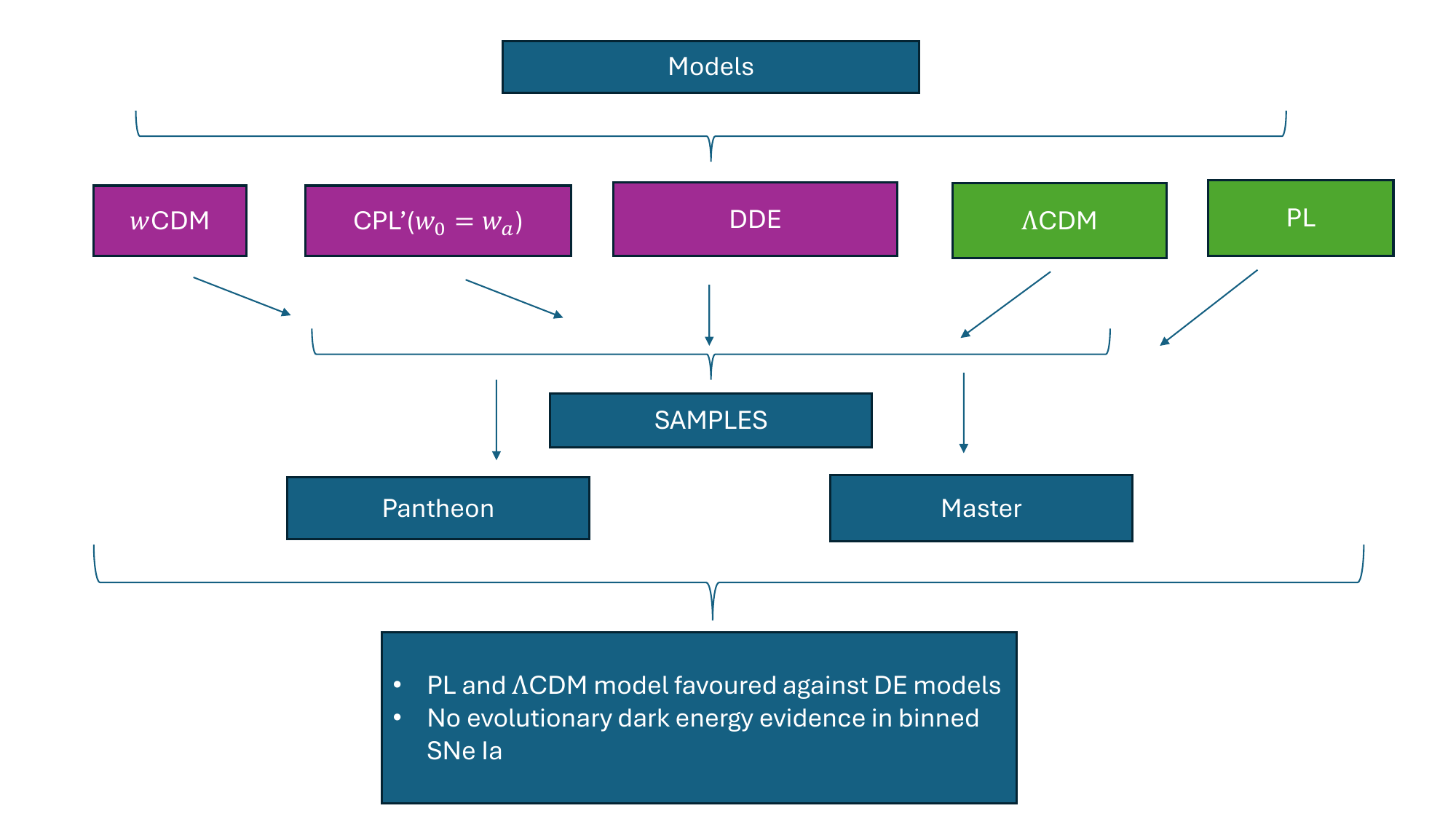}    
\caption{Flow-chart summarizing the structure of the analysis and main results. The chart illustrates the SNe Ia datasets used (Pantheon and Master), the dark energy models tested ($w$CDM, DDE, and CPL' with $w_0 = w_a$) against the reference models ($\Lambda$CDM and PL), and the key outcomes. 
}
\label{fig:flowchart}
\end{figure}

A flowchart that summarizes the analysis and results is visible in Figure~\ref{fig:flowchart}.      

\subsection{Background analysis}
\begin{figure}[ht!]
\centering
\includegraphics[width=0.45\textwidth]{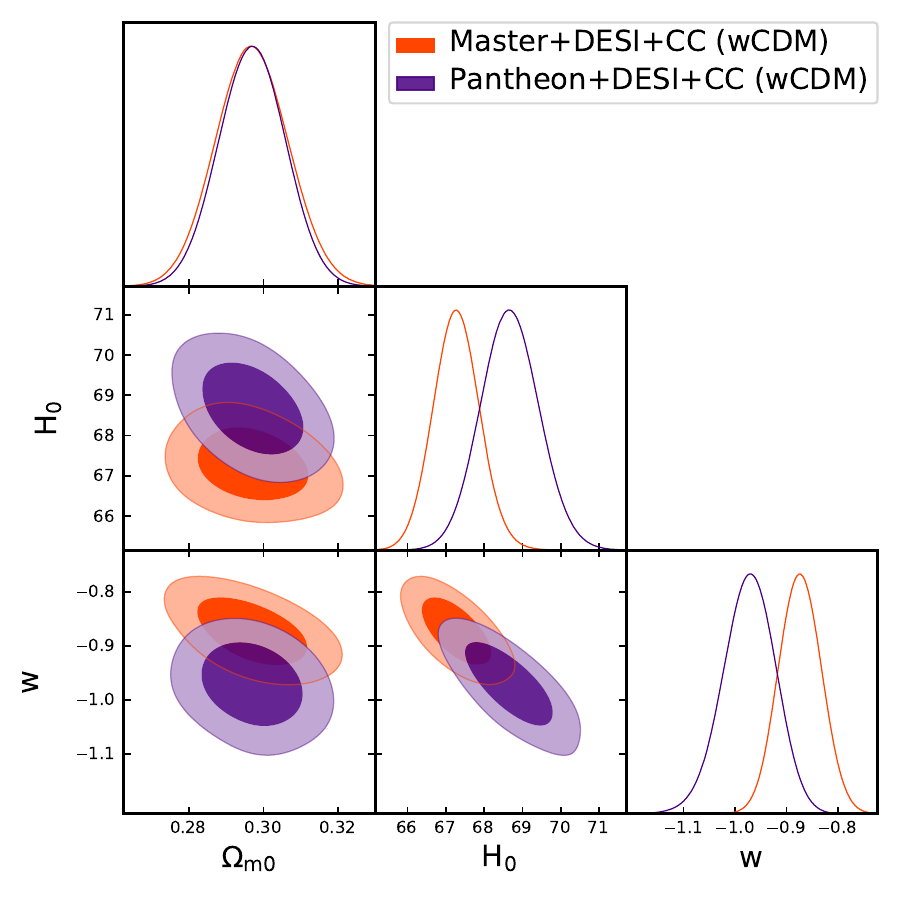}
\includegraphics[width=0.45\textwidth]{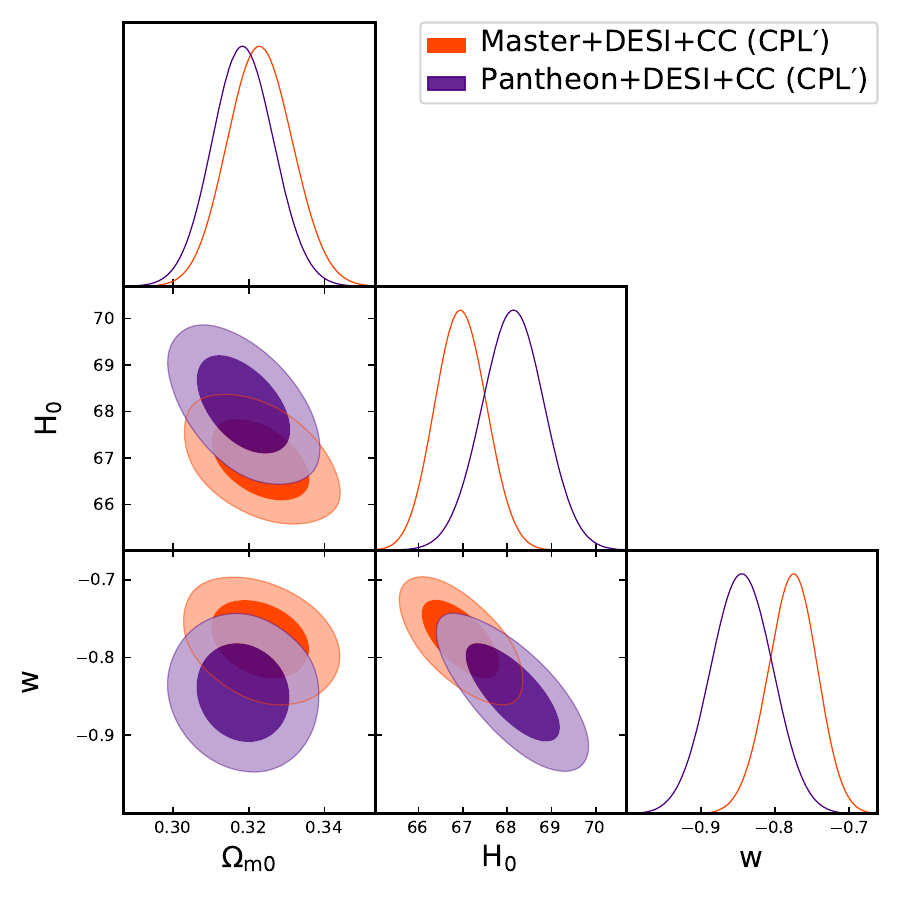}
\includegraphics[width=0.45\textwidth]{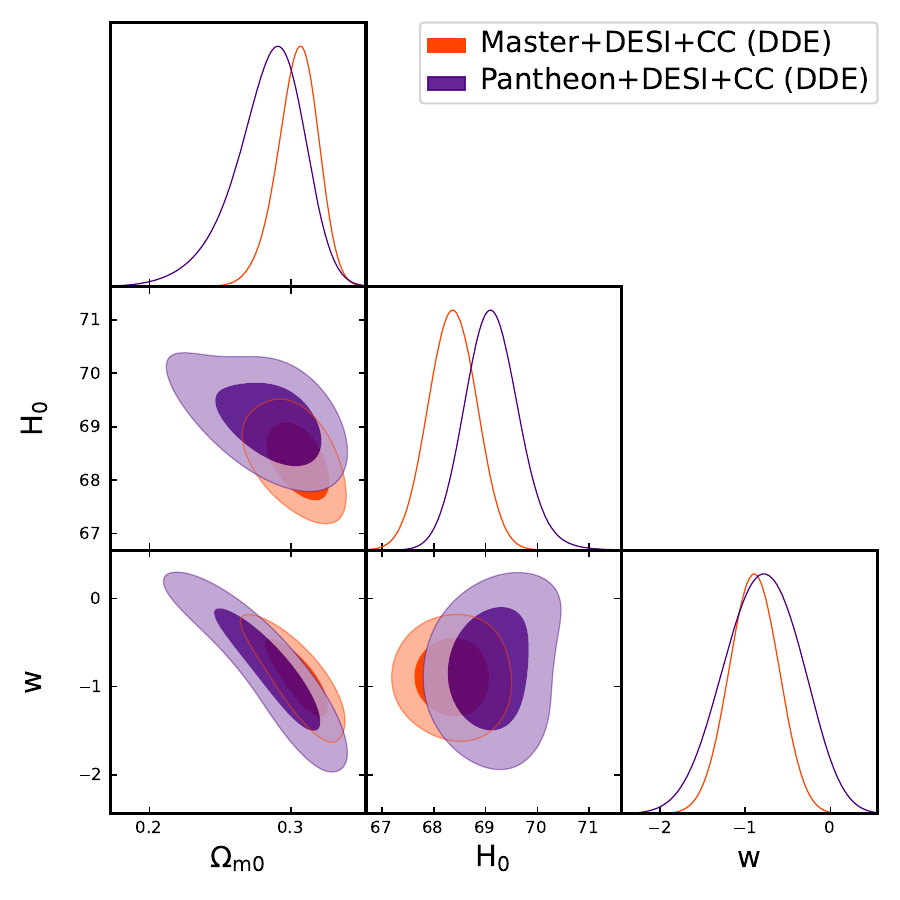}
\includegraphics[width=0.45\textwidth]{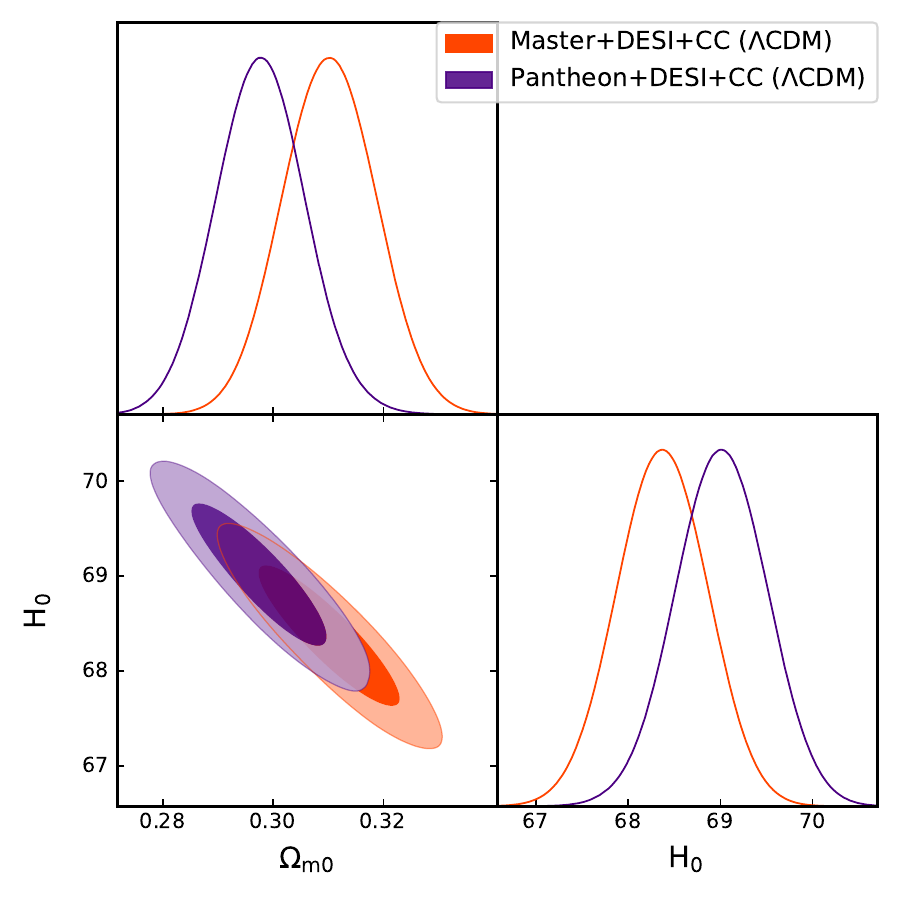}
\caption{One-dimensional posterior probability distributions and two-dimensional $68 \%$ and $95\%$ confidence level contours on cosmological parameters, obtained from the background analysis for the three dark energy models ($w$CDM, CPL' with $w_0 = w_a$, and DDE) and for the $\Lambda$CDM model.}
\label{fig:background_all}
\end{figure}

\noindent Table~\ref{tab:background} presents the results of the standard Background analysis, and specifically parameter constraints obtained by combining SNe Ia (Master or Pantheon), DESI, and CC datasets. We note that, when all background probes are considered together, the analysis does not allow a $1\,\sigma$ discrimination between phantom and quintessence regimes for the $w$CDM and DDE models. In contrast, the CPL' parametrization continues to favor a low-redshift quintessence-like behavior for dark energy, consistent with the results from both the binned and full SNe Ia analyses. However, in this case, the statistical uncertainties associated with the background-only analysis are larger than those obtained from the binned SNe Ia data alone. 
Figure~\ref{fig:background_all} further illustrate the contour plots of the cosmological parameters for the DE models and $\Lambda$CDM model. 

\begin{table}[ht!]
\centering
\scriptsize                          
\setlength{\tabcolsep}{3pt}          
\renewcommand{\arraystretch}{1.6}    
\resizebox{0.8\textwidth}{!}{        
  \begin{tabular}{
    >{\centering\arraybackslash}p{1.5cm}  
    >{\centering\arraybackslash}p{1.5cm}
    c c
  }
  \hline
    &  & \multicolumn{2}{c}{Datasets} \\
  \cline{3-4}
    &  & \textbf{Master+DESI+CC} & \textbf{Pantheon+DESI+CC} \\
  \hline\hline
  \multirow{3}{*}{\parbox[c]{1.5cm}{\centering\rotatebox{90}{$w$CDM}}} 
    & \parbox[c]{1.5cm}{\centering{$H_0$}} 
      & $67.27\pm 0.57 $ & $68.67\pm 0.72$ \\
  \cline{2-4}
    & \parbox[c]{1.5cm}{\centering{$\Omega_{m0}$}} 
      & $0.2969\pm 0.0092 $ & $0.2970\pm 0.0084$ \\
  \cline{2-4}
    & \parbox[c]{1.5cm}{\centering{$w$}} 
      & $-0.873\pm 0.039$ & $-0.972\pm 0.048$ \\
  \hline\hline

  \multirow{3}{*}{\parbox[c]{1.5cm}{\centering\rotatebox{90}{CPL$'$}}}
    & \parbox[c]{1.5cm}{\centering{$H_0$}} 
      & $66.95\pm 0.55 $ & $68.14\pm 0.66$ \\
  \cline{2-4}
    & \parbox[c]{1.5cm}{\centering{$\Omega_{m0}$}} 
      & $ 0.3230\pm 0.0080$ & $0.3185\pm 0.0077$ \\
  \cline{2-4}
 & \parbox[c]{1.5cm}{\centering{$w$}} 
      & $-0.777\pm 0.032 $ & $-0.845\pm 0.039$ \\
  \hline
  \hline
  \multirow{3}{*}{\parbox[c]{1.5cm}{\centering\rotatebox{90}{DDE}}} 
    & \parbox[c]{1.5cm}{\centering{$H_0$}} 
      & $68.36\pm 0.45 $ & $69.12\pm 0.51
 $ \\
  \cline{2-4}
    & \parbox[c]{1.5cm}{\centering{$\Omega_{m0}$}} 
      & $0.305^{+0.015}_{-0.013} $ & $0.284^{+0.028}_{-0.020} $ \\
  \cline{2-4}
    & \parbox[c]{1.5cm}{\centering{$w$}} 
      & $-0.90\pm 0.28$ & $-0.79\pm 0.43$ \\
  \hline
  \hline
  \multirow{2}{*}{\parbox[c]{1.5cm}{\centering\rotatebox{90}{\tiny $\Lambda$CDM}}} 
    & \parbox[c]{1.5cm}{\centering{$H_0$}} 
      & $68.37\pm 0.46 $ & $69.01\pm 0.47$ \\
  \cline{2-4}
    & \parbox[c]{1.5cm}{\centering{$\Omega_{m0}$}} 
      & $0.3102\pm 0.0079 $ & $0.2977\pm 0.0077$ \\
  \hline
  \end{tabular}
}
\caption{Mean values and associated uncertainties of the inferred parameters from the
MCMC background analysis for the three models, i.e. $w$CDM, the reduced CPL model (CPL' with $w_0=w_a$), and the theoretical DDE, are presented for both the binned SNe data samples and full SNe datasets.}
\label{tab:background}
\end{table}

\subsection{Reconstructed effective running Hubble constant and equations of state}

\begin{figure}[ht!]
  \centering
  \includegraphics[width=0.9\linewidth]{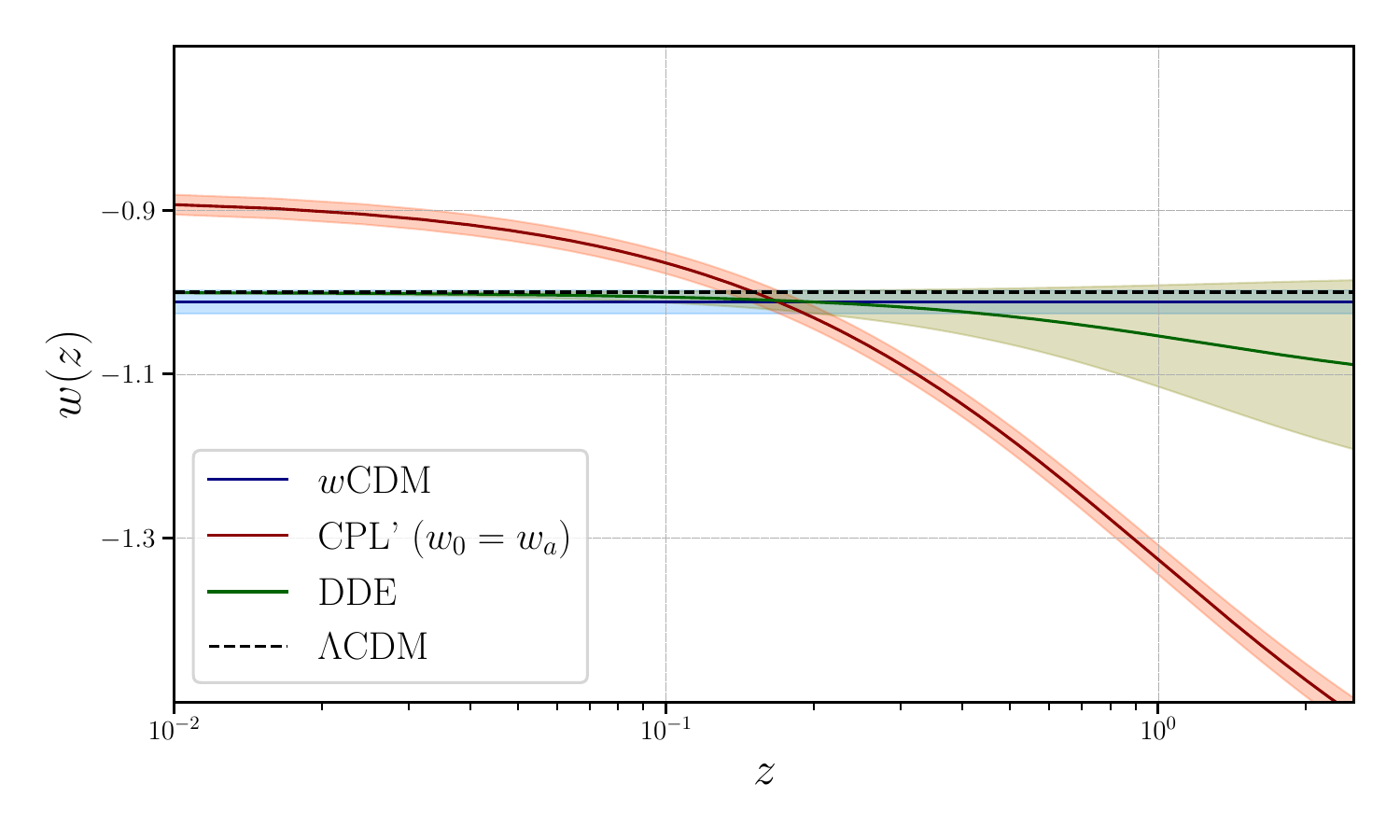}
  \caption{Evolution of the EoS $w(z)$ for the three evolutionary dark energy models. The central bold lines correspond to the mean values form  Table~\ref{tab:binned_full} of Master binned sample and the bands represent the $1\,\sigma$ reconstructed region of probability. The black dashed line stands for the cosmological constant case $w=-1$.}
  \label{fig:EoS}
\end{figure}

\noindent In Figure~\ref{fig:EoS}, the evolution of the DE equation of state, $w(z)$, is presented for all three models, clearly illustrating that $w$CDM and DDE models predict a phantom behavior for DE at low-to-intermediate redshifts while CPL' case results in a DE initially quintessence that becomes phantom at a crossing redshift  $z_c \sim 0.2$.

\begin{figure}[ht!]
\centering
\includegraphics[width=0.48\textwidth]{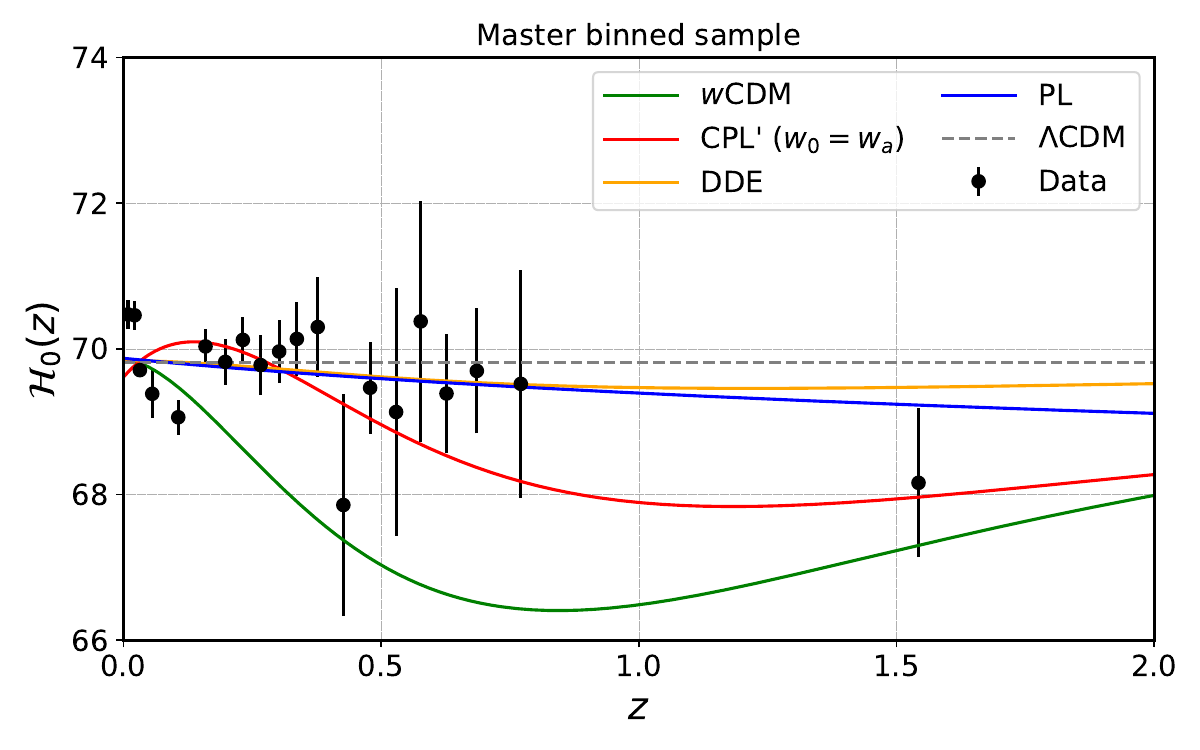}
\includegraphics[width=0.48\textwidth]{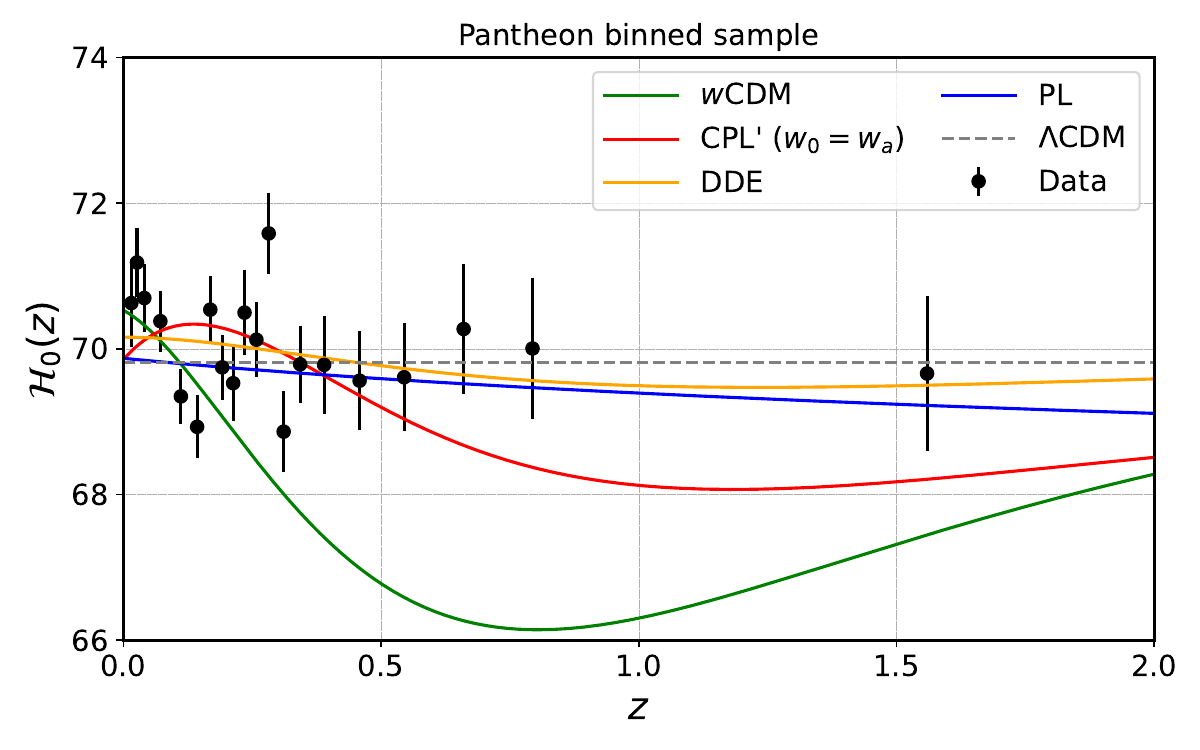}
\caption{Effective running Hubble constant reproduced for the binned Master (left) and Pantheon (right) with the mean values of the fitted parameters in Table~\ref{tab:binned_full}.}
\label{fig:H0z_rec}
\end{figure}

Complementarily, Figure~\ref{fig:H0z_rec} depicts the effective running Hubble constant $\mathcal{H}_0(z)$ as reconstructed for all the models analyzed using the two binned samples, i.e. Master and Pantheon. These reconstructions underline significant deviations from the constant $\Lambda$CDM model, particularly emphasizing the decreasing trend characteristic of the power-law model.

\section{Summary and conclusions}\label{sec6}

\noindent We started by analyzing the theoretical expression of the effective running Hubble constant, as it was introduced in \citet{dainotti2021} and in \citet{schiavone2024}. Actually, we evaluated $\mathcal{H}_0(z)$ in 
correspondence to a $w$CDM and a $w_0w_a\mathrm{CDM}$ models, outlining how the increasing or decreasing behavior in $z$ of this quantity are markers either of a quintessence or of a phantom nature of the dark energy, respectively.
In fact, we have shown in Figure~\ref{fig:H0z_two}, that, for $w(z)>-1$ the derivative $d\mathcal{H}_0/dz$ is always positive, while becoming negative when $w(z)<-1$, the case $w(z)=1$ (i.e., a $\Lambda$CDM-model) clearly implies, as it must, a uniform effective running Hubble constant within the studied redshift range. 

Based on this theoretical assessment, we studied here three different models: 
i) $w$CDM-model ; ii) a specific CPL model with $w_0=w_a$ called CPL'; iii) a dynamics characterized by a process of dark energy creation by the varying cosmological field (DDE model). 
The $w$CDM and DDE imply a monotonic (either increasing or decreasing) behavior of $\mathcal{H}_0(z)$ in the low-redshift regime, whereas the CPL' model allows for a local maximum or minimum of $\mathcal{H}_0(z)$, in correspondence to a given redshift value $z_c\sim 0.2$.

We then tested the associated Hubble parameter of these three models with the Pantheon \citep{scolnic2018} and Master \citep{dainotti_Master} binned samples of SNe Ia.  
In the case of the Pantheon sample,  divided into 20 equipopulated bins, according to the analyses developed in \citet{dainotti2021} and in \citet{dainotti_Master}, respectively, we obtained that we are able to discriminate at least in  $1 \, \sigma$ the nature of the dark energy in favor of its phantom character. By means of a MCMC fitting procedure, which allows the variation of all the free model parameters, we are able to compare Eq.\eqref{eq:H_0_z_2} to data and we are always able to determine, within 1 $\sigma$, the emerging phantom nature of the evolutionary dark energy contribution. 

In the CPL' model, the best-fit reconstruction of $\mathcal{H}_0(z)$ exhibits a maximum around $z_c \simeq 0.2$, suggesting a transition in the nature of dark energy: from a quintessence-like behavior at low redshift ($z < z_c$) to a phantom-like regime at higher redshifts. However, despite this feature, the model is strongly disfavored compared to the other models, as indicated by its large $\ln B_{i,ref} > 5$ (see Table ~\ref{tab:statistics}).

For both the Master and the Pantheon samples, the PL is the most favored model among all, showing the lowest Bayesian factor, followed by the $\Lambda$CDM model. This model, in future works, can be considered a particular phenomenological dark energy model, associated to a monotonically decreasing trend of $\mathcal{H}_0(z)$. 
Nevertheless, we can claim that the binned SNe Ia data for both Pantheon and Master seem to significantly prefer a fit based on the evolutionary phantom matter scenario among the dark energy models. \\
The PL model is not an evolution dark energy model, but it reveals the decreasing trend of $\mathcal{H}_0(z)$ and hence shows the existence of a running Hubble constant as a marker of a deviation from the $\Lambda$CDM model. Anyway, our analysis clarified how the diagnostic tool, $\mathcal{H}_0(z)$, can provide valuable information on the nature of dark energy. Hence, an interesting question emerges about the possibility of representing the PL behavior using a real evolution of the dark energy or other physical effects. For instance, \citet{montani2025_entropy} shows that a model involving interaction between dark energy and dark matter provides a viable framework to reproduce the monotonically decreasing trend of the phenomenological PL behavior, which mimics the observed trend of $H_0$ with the redshift. 

Given that this is the best fit to the data, we are prone to infer that the SNe Ia data, at this current stage, seem unable to produce a quintessence phantom matter transition like the one shown by DESI.

Thus, the present study has the merit of outlining three basic points, as follows: \\
\begin{enumerate}
    \item the theoretical expression of the effective running Hubble constant is able to discriminate between $w(z)>-1$ and $w(z)<-1$ simply looking at its increasing or decreasing behavior with the redshift, respectively; \\
    \item when fitting the Pantheon binned data to the effective running Hubble constant, we are able to determine, at least in $1\, \sigma$ the nature of the evolutionary dark energy; \\
    \item the binned SNe Ia data suggest that the whole Universe evolution is characterized, for low $z$-values, by a real or effective phantom matter component.
\end{enumerate}
    
\noindent These results, however, are subjected to the intrinsic limitation of the statistics in the binned analysis.
On the one hand, these results may suggest that more statistics in the incoming data could favor a very late quintessence era, here not outlined, while, on the other hand, it appears anyway not easy to reconcile the SNe Ia predictions with the ones of DESI, according to which near $z\simeq 0.3$ the transition should take place, i.e. in a redshift region where the binned data appear associated to a decreasing $\mathcal{H}_0(z)$. 

\paragraph{\textbf{Data Availability Statement}}
Data supporting the findings of this study are available upon request for the Master sample.

\paragraph{\textbf{Code Availability Statement}}
The code used for the findings of this study is available upon request.

\paragraph{\textbf{CRediT statement}} 
E. Fazzari: investigation; conceptualization; data analysis and writing – original draft; M. G. Dainotti: data analysis; writing – original draft and supervision; G. Montani: conceptualization; writing – original draft and supervision; A. Melchiorri: writing – review and contribution to discussion.

\section*{Acknowledgements}
\noindent G. Montani would like to thank Tiziano Schiavone for interesting discussions and suggestions on the implementation of particle creation to the evolutionary dark energy paradigm. Authors E. Fazzari and A. Melchiorri are supported by "Theoretical Astroparticle Physics" (TAsP), iniziativa specifica INFN. We acknowledge the use of the wgcosmo GitHub repository by W. Giarè for the Bayes factor computation. The work of E.F. and A.M. was partially supported by the research grant number 2022E2J4RK “PANTHEON: Perspectives in Astroparticle and Neutrino THEory with Old and New messengers” under the program PRIN 2022 funded by the Italian Ministero dell’Università e della Ricerca (MUR). E. F. acknowledges the IT Services at The University of Sheffield for providing High Performance Computing resources. M.G.D. acknowledges the support of the DoS and by JSPS Grant-in-Aid for Scientific Research (KAKENHI) (A), Grant Number JP25H00675. \\

\appendix

\section{Impact of varying \texorpdfstring{$\Omega_{m0}$}{Omega\_m0} in the \texorpdfstring{$\mathcal{H}_0(z)$}{H0(z)} function}\label{sec:appA}

\begin{figure}[H]
\centering
\includegraphics[width=0.48\textwidth]
{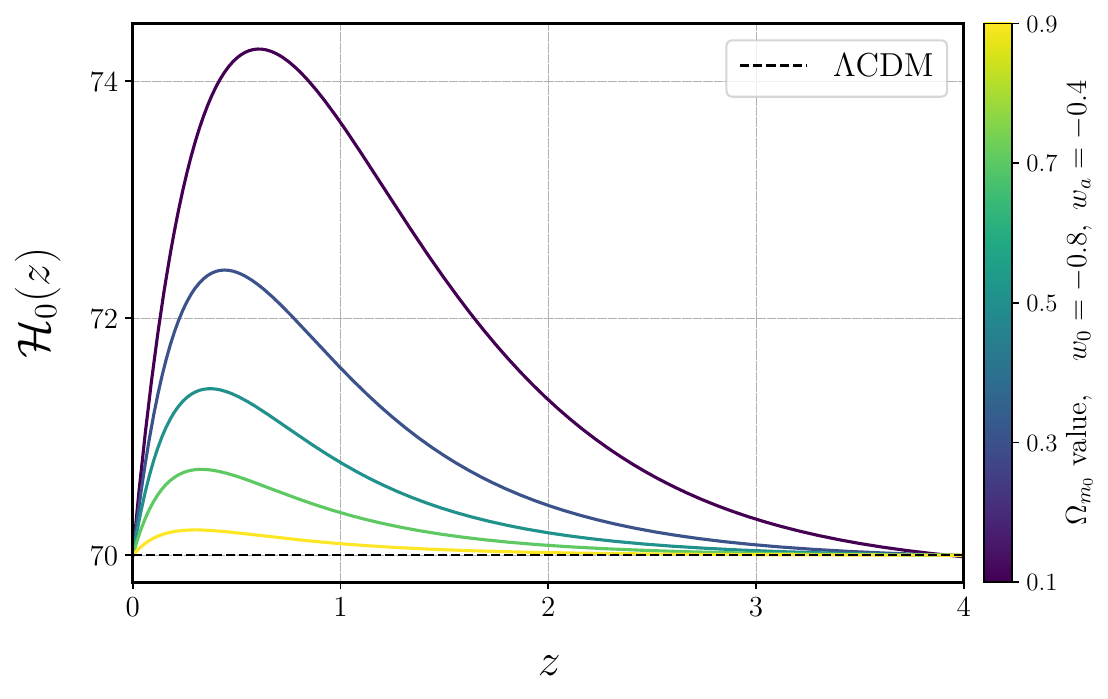}
\includegraphics[width=0.48\textwidth]{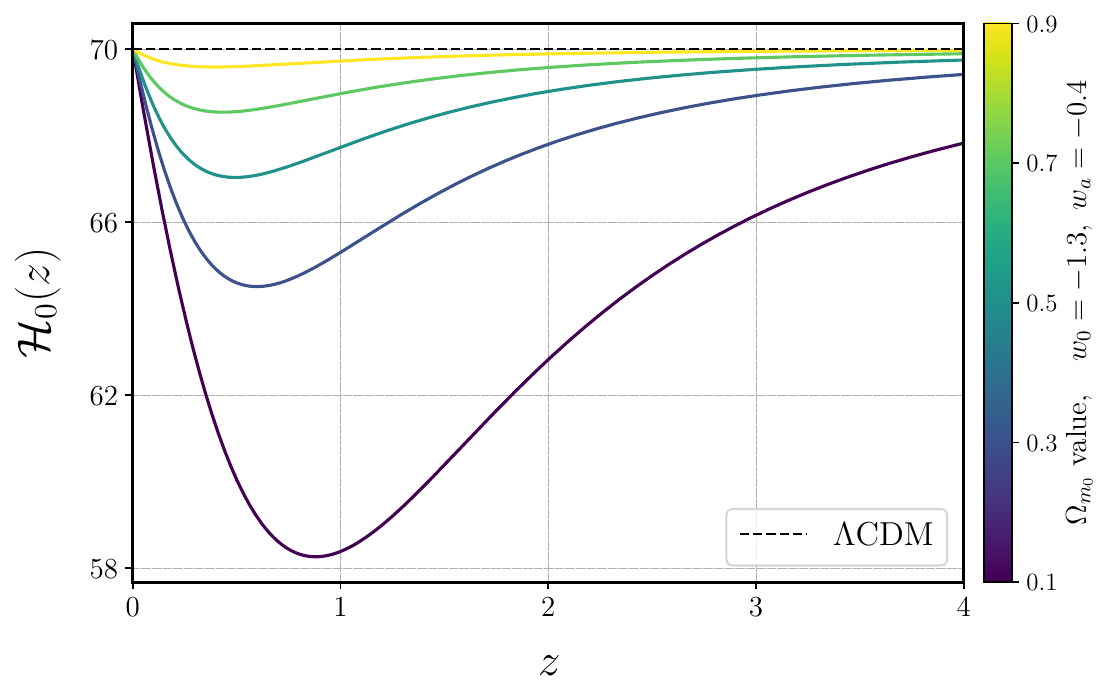}
\caption{Effect of varying the matter density parameter $\Omega_{m0}$ on the effective running Hubble constant. Left panel shows the quintessence case ($w_0=-0.8$, $w_a=-0.4$) while the right panel displays a phantom scenario ($w_0=-1.3$, $w_a=-0.4$) for DE.}
\label{fig:H0z_Omegam}
\end{figure}

\noindent To illustrate the impact of varying the matter density, Figure~\ref{fig:H0z_Omegam} shows $\mathcal{H}_0(z)$ for fixed $H_0$, $w_0$, and $w_a$, while varying $\Omega_{m0}$.

\begin{thebibliography}{73}
\expandafter\ifx\csname natexlab\endcsname\relax\def\natexlab#1{#1}\fi
\providecommand{\url}[1]{\texttt{#1}}
\providecommand{\href}[2]{#2}
\providecommand{\path}[1]{#1}
\providecommand{\DOIprefix}{doi:}
\providecommand{\ArXivprefix}{arXiv:}
\providecommand{\URLprefix}{URL: }
\providecommand{\Pubmedprefix}{pmid:}
\providecommand{\doi}[1]{\href{http://dx.doi.org/#1}{\path{#1}}}
\providecommand{\Pubmed}[1]{\href{pmid:#1}{\path{#1}}}
\providecommand{\bibinfo}[2]{#2}
\ifx\xfnm\relax \def\xfnm[#1]{\unskip,\space#1}\fi
\bibitem[{Abbott et~al.(2025)Abbott, Acevedo, Aguena, Alarcon, Allam, Alves, Amon, Andrade-Oliveira, Annis, Armstrong, Asorey, Avila, Bacon, Bassett, Bechtol, Bernardinelli, Bernstein, Bertin, Blazek, Bocquet, Brooks, Brout, Buckley-Geer, Burke, Camacho, Camilleri, Campos, Rosell, Carollo, Carr, Carretero, Castander, Cawthon, Chang, Chen, Choi, Conselice, Costanzi, da~Costa, Crocce, Davis, DePoy, Desai, Diehl, Dixon, Dodelson, Doel, Doux, Drlica-Wagner, Elvin-Poole, Everett, Ferrero, Ferté, Flaugher, Foley, Fosalba, Friedel, Frieman, Frohmaier, Galbany, García-Bellido, Gatti, Gaztanaga, Giannini, Glazebrook, Graur, Gruen, Gruendl, Gutierrez, Hartley, Herner, Hinton, Hollowood, Honscheid, Huterer, Jain, James, Jeffrey, Kasai, Kelsey, Kent, Kessler, Kim, Kirshner, Kovacs, Kuehn, Lahav, Lee, Lee, Lewis, Li, Lidman, Lin, Malik, Marshall, Martini, Mena-Fernández, Menanteau, Miquel, Mohr, Mould, Muir, Möller, Neilsen, Nichol, Nugent, Ogando, Palmese, Pan, Paterno, Percival, Pereira, Pieres, Malagón, Popovic,
  Porredon, Prat, Qu, Raveri, Rodríguez-Monroy, Romer, Roodman, Rose, Sako, Sanchez, Cid, Schubnell, Scolnic, Sevilla-Noarbe, Shah, Smith, Smith, Soares-Santos, Suchyta, Sullivan, Suntzeff, Swanson, Sánchez, Tarle, Taylor, Thomas, To, Toy, Troxel, Tucker, Tucker, Uddin, Vincenzi, Walker, Weaverdyck, Wechsler, Weller, Wester, Wiseman, Yamamoto, Yuan, Zhang and Zhang}]{DESy5}
\bibinfo{author}{Abbott, D.C.T.M.C.}, \bibinfo{author}{Acevedo, M.}, \bibinfo{author}{Aguena, M.}, \bibinfo{author}{Alarcon, A.}, \bibinfo{author}{Allam, S.}, \bibinfo{author}{Alves, O.}, \bibinfo{author}{Amon, A.}, \bibinfo{author}{Andrade-Oliveira, F.}, \bibinfo{author}{Annis, J.}, \bibinfo{author}{Armstrong, P.}, \bibinfo{author}{Asorey, J.}, \bibinfo{author}{Avila, S.}, \bibinfo{author}{Bacon, D.}, \bibinfo{author}{Bassett, B.A.}, \bibinfo{author}{Bechtol, K.}, \bibinfo{author}{Bernardinelli, P.H.}, \bibinfo{author}{Bernstein, G.M.}, \bibinfo{author}{Bertin, E.}, \bibinfo{author}{Blazek, J.}, \bibinfo{author}{Bocquet, S.}, \bibinfo{author}{Brooks, D.}, \bibinfo{author}{Brout, D.}, \bibinfo{author}{Buckley-Geer, E.}, \bibinfo{author}{Burke, D.L.}, \bibinfo{author}{Camacho, H.}, \bibinfo{author}{Camilleri, R.}, \bibinfo{author}{Campos, A.}, \bibinfo{author}{Rosell, A.C.}, \bibinfo{author}{Carollo, D.}, \bibinfo{author}{Carr, A.}, \bibinfo{author}{Carretero, J.}, \bibinfo{author}{Castander, F.J.},
  \bibinfo{author}{Cawthon, R.}, \bibinfo{author}{Chang, C.}, \bibinfo{author}{Chen, R.}, \bibinfo{author}{Choi, A.}, \bibinfo{author}{Conselice, C.}, \bibinfo{author}{Costanzi, M.}, \bibinfo{author}{da~Costa, L.N.}, \bibinfo{author}{Crocce, M.}, \bibinfo{author}{Davis, T.M.}, \bibinfo{author}{DePoy, D.L.}, \bibinfo{author}{Desai, S.}, \bibinfo{author}{Diehl, H.T.}, \bibinfo{author}{Dixon, M.}, \bibinfo{author}{Dodelson, S.}, \bibinfo{author}{Doel, P.}, \bibinfo{author}{Doux, C.}, \bibinfo{author}{Drlica-Wagner, A.}, \bibinfo{author}{Elvin-Poole, J.}, \bibinfo{author}{Everett, S.}, \bibinfo{author}{Ferrero, I.}, \bibinfo{author}{Ferté, A.}, \bibinfo{author}{Flaugher, B.}, \bibinfo{author}{Foley, R.J.}, \bibinfo{author}{Fosalba, P.}, \bibinfo{author}{Friedel, D.}, \bibinfo{author}{Frieman, J.}, \bibinfo{author}{Frohmaier, C.}, \bibinfo{author}{Galbany, L.}, \bibinfo{author}{García-Bellido, J.}, \bibinfo{author}{Gatti, M.}, \bibinfo{author}{Gaztanaga, E.}, \bibinfo{author}{Giannini, G.},
  \bibinfo{author}{Glazebrook, K.}, \bibinfo{author}{Graur, O.}, \bibinfo{author}{Gruen, D.}, \bibinfo{author}{Gruendl, R.A.}, \bibinfo{author}{Gutierrez, G.}, \bibinfo{author}{Hartley, W.G.}, \bibinfo{author}{Herner, K.}, \bibinfo{author}{Hinton, S.R.}, \bibinfo{author}{Hollowood, D.L.}, \bibinfo{author}{Honscheid, K.}, \bibinfo{author}{Huterer, D.}, \bibinfo{author}{Jain, B.}, \bibinfo{author}{James, D.J.}, \bibinfo{author}{Jeffrey, N.}, \bibinfo{author}{Kasai, E.}, \bibinfo{author}{Kelsey, L.}, \bibinfo{author}{Kent, S.}, \bibinfo{author}{Kessler, R.}, \bibinfo{author}{Kim, A.G.}, \bibinfo{author}{Kirshner, R.P.}, \bibinfo{author}{Kovacs, E.}, \bibinfo{author}{Kuehn, K.}, \bibinfo{author}{Lahav, O.}, \bibinfo{author}{Lee, J.}, \bibinfo{author}{Lee, S.}, \bibinfo{author}{Lewis, G.F.}, \bibinfo{author}{Li, T.S.}, \bibinfo{author}{Lidman, C.}, \bibinfo{author}{Lin, H.}, \bibinfo{author}{Malik, U.}, \bibinfo{author}{Marshall, J.L.}, \bibinfo{author}{Martini, P.}, \bibinfo{author}{Mena-Fernández, J.},
  \bibinfo{author}{Menanteau, F.}, \bibinfo{author}{Miquel, R.}, \bibinfo{author}{Mohr, J.J.}, \bibinfo{author}{Mould, J.}, \bibinfo{author}{Muir, J.}, \bibinfo{author}{Möller, A.}, \bibinfo{author}{Neilsen, E.}, \bibinfo{author}{Nichol, R.C.}, \bibinfo{author}{Nugent, P.}, \bibinfo{author}{Ogando, R.L.C.}, \bibinfo{author}{Palmese, A.}, \bibinfo{author}{Pan, Y.C.}, \bibinfo{author}{Paterno, M.}, \bibinfo{author}{Percival, W.J.}, \bibinfo{author}{Pereira, M.E.S.}, \bibinfo{author}{Pieres, A.}, \bibinfo{author}{Malagón, A.A.P.}, \bibinfo{author}{Popovic, B.}, \bibinfo{author}{Porredon, A.}, \bibinfo{author}{Prat, J.}, \bibinfo{author}{Qu, H.}, \bibinfo{author}{Raveri, M.}, \bibinfo{author}{Rodríguez-Monroy, M.}, \bibinfo{author}{Romer, A.K.}, \bibinfo{author}{Roodman, A.}, \bibinfo{author}{Rose, B.}, \bibinfo{author}{Sako, M.}, \bibinfo{author}{Sanchez, E.}, \bibinfo{author}{Cid, D.S.}, \bibinfo{author}{Schubnell, M.}, \bibinfo{author}{Scolnic, D.}, \bibinfo{author}{Sevilla-Noarbe, I.},
  \bibinfo{author}{Shah, P.}, \bibinfo{author}{Smith, J.A.}, \bibinfo{author}{Smith, M.}, \bibinfo{author}{Soares-Santos, M.}, \bibinfo{author}{Suchyta, E.}, \bibinfo{author}{Sullivan, M.}, \bibinfo{author}{Suntzeff, N.}, \bibinfo{author}{Swanson, M.E.C.}, \bibinfo{author}{Sánchez, B.O.}, \bibinfo{author}{Tarle, G.}, \bibinfo{author}{Taylor, G.}, \bibinfo{author}{Thomas, D.}, \bibinfo{author}{To, C.}, \bibinfo{author}{Toy, M.}, \bibinfo{author}{Troxel, M.A.}, \bibinfo{author}{Tucker, B.E.}, \bibinfo{author}{Tucker, D.L.}, \bibinfo{author}{Uddin, S.A.}, \bibinfo{author}{Vincenzi, M.}, \bibinfo{author}{Walker, A.R.}, \bibinfo{author}{Weaverdyck, N.}, \bibinfo{author}{Wechsler, R.H.}, \bibinfo{author}{Weller, J.}, \bibinfo{author}{Wester, W.}, \bibinfo{author}{Wiseman, P.}, \bibinfo{author}{Yamamoto, M.}, \bibinfo{author}{Yuan, F.}, \bibinfo{author}{Zhang, B.}, \bibinfo{author}{Zhang, Y.}, \bibinfo{year}{2025}.
\newblock \bibinfo{title}{The dark energy survey: Cosmology results with ~1500 new high-redshift type ia supernovae using the full 5-year dataset}.
\newblock \URLprefix \url{https://arxiv.org/abs/2401.02929}, \href{http://arxiv.org/abs/2401.02929}{{\tt arXiv:2401.02929}}.
\bibitem[{Abdul~Karim et~al.(2025)}]{desi2}
\bibinfo{author}{Abdul~Karim, M.}, et~al. (\bibinfo{collaboration}{DESI}), \bibinfo{year}{2025}.
\newblock \bibinfo{title}{{DESI DR2 Results II: Measurements of Baryon Acoustic Oscillations and Cosmological Constraints}} \href{http://arxiv.org/abs/2503.14738}{{\tt arXiv:2503.14738}}.
\bibitem[{Adame et~al.(2024)Adame, Aguilar, Ahlen, Alam, Alexander, Alvarez, Alves, Anand, Andrade, Armengaud et~al.}]{desi}
\bibinfo{author}{Adame, A.}, \bibinfo{author}{Aguilar, J.}, \bibinfo{author}{Ahlen, S.}, \bibinfo{author}{Alam, S.}, \bibinfo{author}{Alexander, D.}, \bibinfo{author}{Alvarez, M.}, \bibinfo{author}{Alves, O.}, \bibinfo{author}{Anand, A.}, \bibinfo{author}{Andrade, U.}, \bibinfo{author}{Armengaud, E.}, et~al. (\bibinfo{collaboration}{DESI}), \bibinfo{year}{2024}.
\newblock \bibinfo{title}{{Desi 2024 VI: Cosmological constraints from the measurements of baryon acoustic oscillations}}.
\newblock \bibinfo{journal}{arXiv preprint arXiv:2404.03002} .
\bibitem[{Aghanim et~al.(2020)Aghanim, Akrami, Ashdown, Aumont, Baccigalupi, Ballardini, Banday, Barreiro, Bartolo, Basak et~al.}]{Planck2018}
\bibinfo{author}{Aghanim, N.}, \bibinfo{author}{Akrami, Y.}, \bibinfo{author}{Ashdown, M.}, \bibinfo{author}{Aumont, J.}, \bibinfo{author}{Baccigalupi, C.}, \bibinfo{author}{Ballardini, M.}, \bibinfo{author}{Banday, A.J.}, \bibinfo{author}{Barreiro, R.}, \bibinfo{author}{Bartolo, N.}, \bibinfo{author}{Basak, S.}, et~al., \bibinfo{year}{2020}.
\newblock \bibinfo{title}{{Planck 2018 results-VI. Cosmological parameters}}.
\newblock \bibinfo{journal}{Astronomy \& Astrophysics} \bibinfo{volume}{641}, \bibinfo{pages}{A6}.
\bibitem[{Bargiacchi et~al.(2023)Bargiacchi, Dainotti, Nagataki and Capozziello}]{bargiacchi2023GRB}
\bibinfo{author}{Bargiacchi, G.}, \bibinfo{author}{Dainotti, M.}, \bibinfo{author}{Nagataki, S.}, \bibinfo{author}{Capozziello, S.}, \bibinfo{year}{2023}.
\newblock \bibinfo{title}{{Gamma-ray bursts, quasars, baryonic acoustic oscillations, and supernovae Ia: New statistical insights and cosmological constraints}}.
\newblock \bibinfo{journal}{Monthly Notices of the Royal Astronomical Society} \bibinfo{volume}{521}, \bibinfo{pages}{3909--3924}.
\bibitem[{Betoule et~al.(2014)Betoule, Kessler, Guy, Mosher, Hardin, Biswas, Astier, El-Hage, Konig, Kuhlmann et~al.}]{JLA}
\bibinfo{author}{Betoule, M.}, \bibinfo{author}{Kessler, R.}, \bibinfo{author}{Guy, J.}, \bibinfo{author}{Mosher, J.}, \bibinfo{author}{Hardin, D.}, \bibinfo{author}{Biswas, R.}, \bibinfo{author}{Astier, P.}, \bibinfo{author}{El-Hage, P.}, \bibinfo{author}{Konig, M.}, \bibinfo{author}{Kuhlmann, S.}, et~al., \bibinfo{year}{2014}.
\newblock \bibinfo{title}{{Improved cosmological constraints from a joint analysis of the SDSS-II and SNLS supernova samples}}.
\newblock \bibinfo{journal}{Astronomy \& Astrophysics} \bibinfo{volume}{568}, \bibinfo{pages}{A22}.
\bibitem[{Borghi et~al.(2022)Borghi, Moresco and Cimatti}]{CC_borghi}
\bibinfo{author}{Borghi, N.}, \bibinfo{author}{Moresco, M.}, \bibinfo{author}{Cimatti, A.}, \bibinfo{year}{2022}.
\newblock \bibinfo{title}{{Toward a better understanding of cosmic chronometers: a new measurement of $H(z)$ at $z\sim0.7$}}.
\newblock \bibinfo{journal}{The Astrophysical Journal Letters} \bibinfo{volume}{928}, \bibinfo{pages}{L4}.
\bibitem[{Brout et~al.(2022)Brout, Scolnic, Popovic, Riess, Carr, Zuntz, Kessler, Davis, Hinton, Jones, Kenworthy, Peterson, Said, Taylor, Ali, Armstrong, Charvu, Dwomoh, Meldorf, Palmese, Qu, Rose, Sanchez, Stubbs, Vincenzi, Wood, Brown, Chen, Chambers, Coulter, Dai, Dimitriadis, Filippenko, Foley, Jha, Kelsey, Kirshner, M\"oller, Muir, Nadathur, Pan, Rest, Rojas-Bravo, Sako, Siebert, Smith, Stahl and Wiseman}]{Brout_2022}
\bibinfo{author}{Brout, D.}, \bibinfo{author}{Scolnic, D.}, \bibinfo{author}{Popovic, B.}, \bibinfo{author}{Riess, A.G.}, \bibinfo{author}{Carr, A.}, \bibinfo{author}{Zuntz, J.}, \bibinfo{author}{Kessler, R.}, \bibinfo{author}{Davis, T.M.}, \bibinfo{author}{Hinton, S.}, \bibinfo{author}{Jones, D.}, \bibinfo{author}{Kenworthy, W.D.}, \bibinfo{author}{Peterson, E.R.}, \bibinfo{author}{Said, K.}, \bibinfo{author}{Taylor, G.}, \bibinfo{author}{Ali, N.}, \bibinfo{author}{Armstrong, P.}, \bibinfo{author}{Charvu, P.}, \bibinfo{author}{Dwomoh, A.}, \bibinfo{author}{Meldorf, C.}, \bibinfo{author}{Palmese, A.}, \bibinfo{author}{Qu, H.}, \bibinfo{author}{Rose, B.M.}, \bibinfo{author}{Sanchez, B.}, \bibinfo{author}{Stubbs, C.W.}, \bibinfo{author}{Vincenzi, M.}, \bibinfo{author}{Wood, C.M.}, \bibinfo{author}{Brown, P.J.}, \bibinfo{author}{Chen, R.}, \bibinfo{author}{Chambers, K.}, \bibinfo{author}{Coulter, D.A.}, \bibinfo{author}{Dai, M.}, \bibinfo{author}{Dimitriadis, G.}, \bibinfo{author}{Filippenko, A.V.},
  \bibinfo{author}{Foley, R.J.}, \bibinfo{author}{Jha, S.W.}, \bibinfo{author}{Kelsey, L.}, \bibinfo{author}{Kirshner, R.P.}, \bibinfo{author}{M\"oller, A.}, \bibinfo{author}{Muir, J.}, \bibinfo{author}{Nadathur, S.}, \bibinfo{author}{Pan, Y.C.}, \bibinfo{author}{Rest, A.}, \bibinfo{author}{Rojas-Bravo, C.}, \bibinfo{author}{Sako, M.}, \bibinfo{author}{Siebert, M.R.}, \bibinfo{author}{Smith, M.}, \bibinfo{author}{Stahl, B.E.}, \bibinfo{author}{Wiseman, P.}, \bibinfo{year}{2022}.
\newblock \bibinfo{title}{{The Pantheon+ Analysis: Cosmological Constraints}}.
\newblock \bibinfo{journal}{The Astrophysical Journal} \bibinfo{volume}{938}, \bibinfo{pages}{110}.
\newblock \DOIprefix\doi{10.3847/1538-4357/ac8e04}.
\bibitem[{Calvao et~al.(1992)Calvao, Lima and Waga}]{matcre_calvaoLima}
\bibinfo{author}{Calvao, M.}, \bibinfo{author}{Lima, J.}, \bibinfo{author}{Waga, I.}, \bibinfo{year}{1992}.
\newblock \bibinfo{title}{{On the thermodynamics of matter creation in cosmology}}.
\newblock \bibinfo{journal}{Physics Letters A} \bibinfo{volume}{162}, \bibinfo{pages}{223--226}.
\bibitem[{Chevallier and Polarski(2001)}]{CPL1}
\bibinfo{author}{Chevallier, M.}, \bibinfo{author}{Polarski, D.}, \bibinfo{year}{2001}.
\newblock \bibinfo{title}{{Accelerating universes with scaling dark matter}}.
\newblock \bibinfo{journal}{International Journal of Modern Physics D} \bibinfo{volume}{10}, \bibinfo{pages}{213--223}.
\bibitem[{Colg{\'a}in et~al.(2024)Colg{\'a}in, Dainotti, Capozziello, Pourojaghi, Sheikh-Jabbari and Stojkovic}]{colgain_Dainotti}
\bibinfo{author}{Colg{\'a}in, E.{\'O}.}, \bibinfo{author}{Dainotti, M.G.}, \bibinfo{author}{Capozziello, S.}, \bibinfo{author}{Pourojaghi, S.}, \bibinfo{author}{Sheikh-Jabbari, M.}, \bibinfo{author}{Stojkovic, D.}, \bibinfo{year}{2024}.
\newblock \bibinfo{title}{{Does DESI 2024 Confirm $\Lambda$CDM?}}
\newblock \bibinfo{journal}{arXiv preprint arXiv:2404.08633} .
\bibitem[{Dainotti et~al.(2023a)Dainotti, De~Simone, Montani, Schiavone and Lambiase}]{dainotti_montani2023}
\bibinfo{author}{Dainotti, M.}, \bibinfo{author}{De~Simone, B.}, \bibinfo{author}{Montani, G.}, \bibinfo{author}{Schiavone, T.}, \bibinfo{author}{Lambiase, G.}, \bibinfo{year}{2023}a.
\newblock \bibinfo{title}{{The Hubble constant tension: current status and future perspectives through new cosmological probes}}.
\newblock \bibinfo{journal}{arXiv preprint arXiv:2301.10572} .
\bibitem[{Dainotti et~al.(2023b)Dainotti, Bargiacchi, Bogdan, Lenart, Iwasaki, Capozziello, Zhang and Fraija}]{dainotti_GRB_23}
\bibinfo{author}{Dainotti, M.G.}, \bibinfo{author}{Bargiacchi, G.}, \bibinfo{author}{Bogdan, M.}, \bibinfo{author}{Lenart, A.L.}, \bibinfo{author}{Iwasaki, K.}, \bibinfo{author}{Capozziello, S.}, \bibinfo{author}{Zhang, B.}, \bibinfo{author}{Fraija, N.}, \bibinfo{year}{2023}b.
\newblock \bibinfo{title}{{Reducing the uncertainty on the Hubble constant up to 35\% with an improved statistical analysis: different best-fit likelihoods for type Ia supernovae, baryon acoustic oscillations, quasars, and gamma-ray bursts}}.
\newblock \bibinfo{journal}{The Astrophysical Journal} \bibinfo{volume}{951}, \bibinfo{pages}{63}.
\bibitem[{Dainotti et~al.(2024a)Dainotti, Bargiacchi, Lenart and Capozziello}]{dainotti2024quasars}
\bibinfo{author}{Dainotti, M.G.}, \bibinfo{author}{Bargiacchi, G.}, \bibinfo{author}{Lenart, A.{\L}.}, \bibinfo{author}{Capozziello, S.}, \bibinfo{year}{2024}a.
\newblock \bibinfo{title}{{The scavenger hunt for quasar samples to be used as cosmological tools}}.
\newblock \bibinfo{journal}{Galaxies} \bibinfo{volume}{12}, \bibinfo{pages}{4}.
\bibitem[{Dainotti et~al.(2023c)Dainotti, Bargiacchi, Lenart, Nagataki and Capozziello}]{dainotti2023quasars}
\bibinfo{author}{Dainotti, M.G.}, \bibinfo{author}{Bargiacchi, G.}, \bibinfo{author}{Lenart, A.{\L}.}, \bibinfo{author}{Nagataki, S.}, \bibinfo{author}{Capozziello, S.}, \bibinfo{year}{2023}c.
\newblock \bibinfo{title}{{Quasars: Standard Candles up to $z=7.5$ with the Precision of Supernovae Ia}}.
\newblock \bibinfo{journal}{The Astrophysical Journal} \bibinfo{volume}{950}, \bibinfo{pages}{45}.
\bibitem[{Dainotti et~al.(2021a)Dainotti, De~Simone, Schiavone, Montani, Rinaldi and Lambiase}]{dainotti2021}
\bibinfo{author}{Dainotti, M.G.}, \bibinfo{author}{De~Simone, B.}, \bibinfo{author}{Schiavone, T.}, \bibinfo{author}{Montani, G.}, \bibinfo{author}{Rinaldi, E.}, \bibinfo{author}{Lambiase, G.}, \bibinfo{year}{2021}a.
\newblock \bibinfo{title}{{On the Hubble constant tension in the SNe Ia Pantheon sample}}.
\newblock \bibinfo{journal}{The Astrophysical Journal} \bibinfo{volume}{912}, \bibinfo{pages}{150}.
\bibitem[{Dainotti et~al.(2022)Dainotti, De~Simone, Schiavone, Montani, Rinaldi, Lambiase, Bogdan and Ugale}]{dainotti2022}
\bibinfo{author}{Dainotti, M.G.}, \bibinfo{author}{De~Simone, B.}, \bibinfo{author}{Schiavone, T.}, \bibinfo{author}{Montani, G.}, \bibinfo{author}{Rinaldi, E.}, \bibinfo{author}{Lambiase, G.}, \bibinfo{author}{Bogdan, M.}, \bibinfo{author}{Ugale, S.}, \bibinfo{year}{2022}.
\newblock \bibinfo{title}{{On the evolution of the Hubble constant with the SNe Ia pantheon sample and baryon acoustic oscillations: a feasibility study for GRB-cosmology in 2030}}.
\newblock \bibinfo{journal}{Galaxies} \bibinfo{volume}{10}, \bibinfo{pages}{24}.
\bibitem[{Dainotti et~al.(2024b)Dainotti, Lenart, Yengejeh, Chakraborty, Fraija, Di~Valentino and Montani}]{dainotti2024binquasars}
\bibinfo{author}{Dainotti, M.G.}, \bibinfo{author}{Lenart, A.}, \bibinfo{author}{Yengejeh, M.G.}, \bibinfo{author}{Chakraborty, S.}, \bibinfo{author}{Fraija, N.}, \bibinfo{author}{Di~Valentino, E.}, \bibinfo{author}{Montani, G.}, \bibinfo{year}{2024}b.
\newblock \bibinfo{title}{{A new binning method to choose a standard set of Quasars}}.
\newblock \bibinfo{journal}{Physics of the Dark Universe} \bibinfo{volume}{44}, \bibinfo{pages}{101428}.
\bibitem[{Dainotti et~al.(2021b)Dainotti, Petrosian and Bowden}]{dainotti-petrosian}
\bibinfo{author}{Dainotti, M.G.}, \bibinfo{author}{Petrosian, V.}, \bibinfo{author}{Bowden, L.}, \bibinfo{year}{2021}b.
\newblock \bibinfo{title}{{Cosmological evolution of the formation rate of short gamma-ray bursts with and without extended emission}}.
\newblock \bibinfo{journal}{The Astrophysical Journal Letters} \bibinfo{volume}{914}, \bibinfo{pages}{L40}.
\bibitem[{Dainotti et~al.(2025)Dainotti, Simone, Garg, Kohri, Bashyal, Aich, Mondal, Nagataki, Montani, Jareen, Jabir, Khanjani, Bogdan, Fraija, do~E.~S.~Pedreira, Dejrah, Singh, Parakh, Mandal, Jarial, Lambiase and Sarkar}]{dainotti_Master}
\bibinfo{author}{Dainotti, M.G.}, \bibinfo{author}{Simone, B.D.}, \bibinfo{author}{Garg, A.}, \bibinfo{author}{Kohri, K.}, \bibinfo{author}{Bashyal, A.}, \bibinfo{author}{Aich, A.}, \bibinfo{author}{Mondal, A.}, \bibinfo{author}{Nagataki, S.}, \bibinfo{author}{Montani, G.}, \bibinfo{author}{Jareen, T.}, \bibinfo{author}{Jabir, V.M.}, \bibinfo{author}{Khanjani, S.}, \bibinfo{author}{Bogdan, M.}, \bibinfo{author}{Fraija, N.}, \bibinfo{author}{do~E.~S.~Pedreira, A.C.C.}, \bibinfo{author}{Dejrah, R.H.}, \bibinfo{author}{Singh, A.}, \bibinfo{author}{Parakh, M.}, \bibinfo{author}{Mandal, R.}, \bibinfo{author}{Jarial, K.}, \bibinfo{author}{Lambiase, G.}, \bibinfo{author}{Sarkar, H.}, \bibinfo{year}{2025}.
\newblock \bibinfo{title}{{A New Master Supernovae Ia sample and the investigation of the $H_0$ tension}}.
\newblock \URLprefix \url{https://arxiv.org/abs/2501.11772}, \href{http://arxiv.org/abs/2501.11772}{{\tt arXiv:2501.11772}}.
\bibitem[{De~Simone et~al.(2025)De~Simone, van Putten, Dainotti and Lambiase}]{DeSimone:2024lvy}
\bibinfo{author}{De~Simone, B.}, \bibinfo{author}{van Putten, M.H.P.M.}, \bibinfo{author}{Dainotti, M.G.}, \bibinfo{author}{Lambiase, G.}, \bibinfo{year}{2025}.
\newblock \bibinfo{title}{{A doublet of cosmological models to challenge the H0 tension in the Pantheon Supernovae Ia catalog}}.
\newblock \bibinfo{journal}{JHEAp} \bibinfo{volume}{45}, \bibinfo{pages}{290--298}.
\newblock \DOIprefix\doi{10.1016/j.jheap.2024.12.003}, \href{http://arxiv.org/abs/2411.05744}{{\tt arXiv:2411.05744}}.
\bibitem[{Di~Valentino et~al.(2025)Di~Valentino, Levi~Said, Riess, Pollo, Poulin, G{\'o}mez-Valent, Weltman, Palmese, Huang, Carsten et~al.}]{whitepaper_cosmoverse}
\bibinfo{author}{Di~Valentino, E.}, \bibinfo{author}{Levi~Said, J.}, \bibinfo{author}{Riess, A.}, \bibinfo{author}{Pollo, A.}, \bibinfo{author}{Poulin, V.}, \bibinfo{author}{G{\'o}mez-Valent, A.}, \bibinfo{author}{Weltman, A.}, \bibinfo{author}{Palmese, A.}, \bibinfo{author}{Huang, C.D.}, \bibinfo{author}{Carsten, v.d.B.}, et~al., \bibinfo{year}{2025}.
\newblock \bibinfo{title}{{The CosmoVerse White Paper: Addressing observational tensions in cosmology with systematics and fundamental physics}}.
\newblock \bibinfo{journal}{PHYSICS OF THE DARK UNIVERSE} , \bibinfo{pages}{1--416}.
\bibitem[{Di~Valentino et~al.(2020)Di~Valentino, Melchiorri and Silk}]{divalentino_planck}
\bibinfo{author}{Di~Valentino, E.}, \bibinfo{author}{Melchiorri, A.}, \bibinfo{author}{Silk, J.}, \bibinfo{year}{2020}.
\newblock \bibinfo{title}{{Planck evidence for a closed Universe and a possible crisis for cosmology}}.
\newblock \bibinfo{journal}{Nature Astronomy} \bibinfo{volume}{4}, \bibinfo{pages}{196--203}.
\bibitem[{Di~Valentino et~al.(2021)Di~Valentino, Mena, Pan, Visinelli, Yang, Melchiorri, Mota, Riess and Silk}]{divalentino-Hubbletension}
\bibinfo{author}{Di~Valentino, E.}, \bibinfo{author}{Mena, O.}, \bibinfo{author}{Pan, S.}, \bibinfo{author}{Visinelli, L.}, \bibinfo{author}{Yang, W.}, \bibinfo{author}{Melchiorri, A.}, \bibinfo{author}{Mota, D.F.}, \bibinfo{author}{Riess, A.G.}, \bibinfo{author}{Silk, J.}, \bibinfo{year}{2021}.
\newblock \bibinfo{title}{{In the realm of the Hubble tension—a review of solutions}}.
\newblock \bibinfo{journal}{Classical and Quantum Gravity} \bibinfo{volume}{38}, \bibinfo{pages}{153001}.
\newblock \DOIprefix\doi{10.1088/1361-6382/ac086d}.
\bibitem[{Efstathiou(2021)}]{efstathiou2021}
\bibinfo{author}{Efstathiou, G.}, \bibinfo{year}{2021}.
\newblock \bibinfo{title}{{To $H_0$ or not to $H_0$?}}
\newblock \bibinfo{journal}{Monthly Notices of the Royal Astronomical Society} \bibinfo{volume}{505}, \bibinfo{pages}{3866--3872}.
\bibitem[{Efstathiou and Gratton(2020)}]{efstathiou_planck}
\bibinfo{author}{Efstathiou, G.}, \bibinfo{author}{Gratton, S.}, \bibinfo{year}{2020}.
\newblock \bibinfo{title}{{The evidence for a spatially flat Universe}}.
\newblock \bibinfo{journal}{Monthly Notices of the Royal Astronomical Society: Letters} \bibinfo{volume}{496}, \bibinfo{pages}{L91--L95}.
\bibitem[{Elizalde et~al.(2024)Elizalde, Khurshudyan and Odintsov}]{elizalde_odintsov}
\bibinfo{author}{Elizalde, E.}, \bibinfo{author}{Khurshudyan, M.}, \bibinfo{author}{Odintsov, S.D.}, \bibinfo{year}{2024}.
\newblock \bibinfo{title}{{Can we learn from matter creation to solve the $H_0$ tension problem?}}
\newblock \bibinfo{journal}{The European Physical Journal C} \bibinfo{volume}{84}, \bibinfo{pages}{782}.
\bibitem[{Escamilla et~al.(2024)Escamilla, Fiorucci, Montani and Valentino}]{montani_hubbletension_1}
\bibinfo{author}{Escamilla, L.A.}, \bibinfo{author}{Fiorucci, D.}, \bibinfo{author}{Montani, G.}, \bibinfo{author}{Valentino, E.D.}, \bibinfo{year}{2024}.
\newblock \bibinfo{title}{{Exploring the Hubble tension with a late time Modified Gravity scenario}}.
\newblock \href{http://arxiv.org/abs/2408.04354}{{\tt arXiv:2408.04354}}.
\bibitem[{Favale et~al.(2024)Favale, Dainotti, G{\'o}mez-Valent and Migliaccio}]{Favale_uniform}
\bibinfo{author}{Favale, A.}, \bibinfo{author}{Dainotti, M.G.}, \bibinfo{author}{G{\'o}mez-Valent, A.}, \bibinfo{author}{Migliaccio, M.}, \bibinfo{year}{2024}.
\newblock \bibinfo{title}{{Towards a new model-independent calibration of Gamma-Ray Bursts}}.
\newblock \bibinfo{journal}{JHEAp} \bibinfo{volume}{44}, \bibinfo{pages}{323--339}.
\newblock \DOIprefix\doi{10.1016/j.jheap.2024.10.010}, \href{http://arxiv.org/abs/2402.13115}{{\tt arXiv:2402.13115}}.
\bibitem[{Fazzari et~al.(2025)Fazzari, Leo, Montani, Martinelli, Melchiorri and Cañas-Herrera}]{fazzari_deleo}
\bibinfo{author}{Fazzari, E.}, \bibinfo{author}{Leo, C.D.}, \bibinfo{author}{Montani, G.}, \bibinfo{author}{Martinelli, M.}, \bibinfo{author}{Melchiorri, A.}, \bibinfo{author}{Cañas-Herrera, G.}, \bibinfo{year}{2025}.
\newblock \bibinfo{title}{{Investigating $f(R)$-Inflation: background evolution and constraints}}.
\newblock \URLprefix \url{https://arxiv.org/abs/2507.13890}, \href{http://arxiv.org/abs/2507.13890}{{\tt arXiv:2507.13890}}.
\bibitem[{Gelman and Rubin(1992)}]{gelman_rubin}
\bibinfo{author}{Gelman, A.}, \bibinfo{author}{Rubin, D.B.}, \bibinfo{year}{1992}.
\newblock \bibinfo{title}{{Inference from iterative simulation using multiple sequences}}.
\newblock \bibinfo{journal}{Statistical science} \bibinfo{volume}{7}, \bibinfo{pages}{457--472}.
\bibitem[{Giar{\`e}(2024)}]{giare_dynamical}
\bibinfo{author}{Giar{\`e}, W.}, \bibinfo{year}{2024}.
\newblock \bibinfo{title}{{Dynamical Dark Energy Beyond Planck? Constraints from multiple CMB probes, DESI BAO and Type-Ia Supernovae}}.
\newblock \bibinfo{journal}{arXiv preprint arXiv:2409.17074} .
\bibitem[{Giar{\`e} et~al.(2025)Giar{\`e}, Mahassen, Di~Valentino and Pan}]{giare_overviewDDE}
\bibinfo{author}{Giar{\`e}, W.}, \bibinfo{author}{Mahassen, T.}, \bibinfo{author}{Di~Valentino, E.}, \bibinfo{author}{Pan, S.}, \bibinfo{year}{2025}.
\newblock \bibinfo{title}{{An overview of what current data can (and cannot yet) say about evolving dark energy}}.
\newblock \bibinfo{journal}{Physics of the Dark Universe} , \bibinfo{pages}{101906}.
\bibitem[{Giarè(2025)}]{William_git}
\bibinfo{author}{Giarè, W.}, \bibinfo{year}{2025}.
\newblock \bibinfo{title}{{wgcosmo GitHub repository}}.
\newblock \bibinfo{howpublished}{\url{https://github.com/williamgiare/wgcosmo.git}}.
\bibitem[{Giarè et~al.(2024)Giarè, Najafi, Pan, Di~Valentino and Firouzjaee}]{Giare_Robust_DDE}
\bibinfo{author}{Giarè, W.}, \bibinfo{author}{Najafi, M.}, \bibinfo{author}{Pan, S.}, \bibinfo{author}{Di~Valentino, E.}, \bibinfo{author}{Firouzjaee, J.T.}, \bibinfo{year}{2024}.
\newblock \bibinfo{title}{{Robust preference for Dynamical Dark Energy in DESI BAO and SN measurements}}.
\newblock \bibinfo{journal}{Journal of Cosmology and Astroparticle Physics} \bibinfo{volume}{2024}, \bibinfo{pages}{035}.
\newblock \URLprefix \url{http://dx.doi.org/10.1088/1475-7516/2024/10/035}, \DOIprefix\doi{10.1088/1475-7516/2024/10/035}.
\bibitem[{Jeffreys(1998)}]{jeffreys_scale}
\bibinfo{author}{Jeffreys, H.}, \bibinfo{year}{1998}.
\newblock \bibinfo{title}{The theory of probability}.
\newblock \bibinfo{publisher}{OuP Oxford}.
\bibitem[{Jia et~al.(2023)Jia, Hu and Wang}]{cinesi_H0z}
\bibinfo{author}{Jia, X.}, \bibinfo{author}{Hu, J.}, \bibinfo{author}{Wang, F.}, \bibinfo{year}{2023}.
\newblock \bibinfo{title}{{Evidence of a decreasing trend for the Hubble constant}}.
\newblock \bibinfo{journal}{Astronomy \& Astrophysics} \bibinfo{volume}{674}, \bibinfo{pages}{A45}.
\bibitem[{Jiang et~al.(2024)Jiang, Pedrotti, da~Costa and Vagnozzi}]{Jiang:2024xnu}
\bibinfo{author}{Jiang, J.Q.}, \bibinfo{author}{Pedrotti, D.}, \bibinfo{author}{da~Costa, S.S.}, \bibinfo{author}{Vagnozzi, S.}, \bibinfo{year}{2024}.
\newblock \bibinfo{title}{{Nonparametric late-time expansion history reconstruction and implications for the Hubble tension in light of recent DESI and type Ia supernovae data}}.
\newblock \bibinfo{journal}{Phys. Rev. D} \bibinfo{volume}{110}, \bibinfo{pages}{123519}.
\newblock \DOIprefix\doi{10.1103/PhysRevD.110.123519}, \href{http://arxiv.org/abs/2408.02365}{{\tt arXiv:2408.02365}}.
\bibitem[{Jimenez and Loeb(2002)}]{Jimenez_247}
\bibinfo{author}{Jimenez, R.}, \bibinfo{author}{Loeb, A.}, \bibinfo{year}{2002}.
\newblock \bibinfo{title}{{Constraining Cosmological Parameters Based on Relative Galaxy Ages}}.
\newblock \bibinfo{journal}{The Astrophysical Journal} \bibinfo{volume}{573}, \bibinfo{pages}{37–42}.
\newblock \DOIprefix\doi{10.1086/340549}.
\bibitem[{Kazantzidis and Perivolaropoulos(2020)}]{kazantzidis2020}
\bibinfo{author}{Kazantzidis, L.}, \bibinfo{author}{Perivolaropoulos, L.}, \bibinfo{year}{2020}.
\newblock \bibinfo{title}{{Hints of a local matter underdensity or modified gravity in the low z Pantheon data}}.
\newblock \bibinfo{journal}{Physical Review D} \bibinfo{volume}{102}, \bibinfo{pages}{023520}.
\bibitem[{Krishnan and Mondol(2022)}]{krishnan2022}
\bibinfo{author}{Krishnan, C.}, \bibinfo{author}{Mondol, R.}, \bibinfo{year}{2022}.
\newblock \bibinfo{title}{{$H_0 $ as a Universal FLRW Diagnostic}}.
\newblock \bibinfo{journal}{arXiv preprint arXiv:2201.13384} .
\bibitem[{Krishnan et~al.(2021)Krishnan, {\'O}~Colg{\'a}in, Sheikh-Jabbari and Yang}]{krishnan2021}
\bibinfo{author}{Krishnan, C.}, \bibinfo{author}{{\'O}~Colg{\'a}in, E.}, \bibinfo{author}{Sheikh-Jabbari, M.}, \bibinfo{author}{Yang, T.}, \bibinfo{year}{2021}.
\newblock \bibinfo{title}{{Running Hubble tension and a H0 diagnostic}}.
\newblock \bibinfo{journal}{Physical Review D} \bibinfo{volume}{103}, \bibinfo{pages}{103509}.
\bibitem[{Lenart et~al.(2023)Lenart, Bargiacchi, Dainotti, Nagataki and Capozziello}]{lenart2023}
\bibinfo{author}{Lenart, A.{\L}.}, \bibinfo{author}{Bargiacchi, G.}, \bibinfo{author}{Dainotti, M.G.}, \bibinfo{author}{Nagataki, S.}, \bibinfo{author}{Capozziello, S.}, \bibinfo{year}{2023}.
\newblock \bibinfo{title}{{A bias-free cosmological analysis with quasars alleviating $H_0$ tension}}.
\newblock \bibinfo{journal}{The Astrophysical Journal Supplement Series} \bibinfo{volume}{264}, \bibinfo{pages}{46}.
\bibitem[{Lewis(2019)}]{getdist}
\bibinfo{author}{Lewis, A.}, \bibinfo{year}{2019}.
\newblock \bibinfo{title}{{GetDist: a Python package for analysing Monte Carlo samples}}.
\newblock \bibinfo{journal}{arXiv preprint arXiv:1910.13970} .
\bibitem[{Linder(2003)}]{CPL2}
\bibinfo{author}{Linder, E.V.}, \bibinfo{year}{2003}.
\newblock \bibinfo{title}{{Exploring the expansion history of the universe}}.
\newblock \bibinfo{journal}{Physical review letters} \bibinfo{volume}{90}, \bibinfo{pages}{091301}.
\bibitem[{Linder(2024)}]{linder2024}
\bibinfo{author}{Linder, E.V.}, \bibinfo{year}{2024}.
\newblock \bibinfo{title}{Interpreting dark energy data away from $\lambda$}.
\newblock \URLprefix \url{https://arxiv.org/abs/2410.10981}, \href{http://arxiv.org/abs/2410.10981}{{\tt arXiv:2410.10981}}.
\bibitem[{Lodha et~al.(2025)Lodha, Calderon, Matthewson, Shafieloo, Ishak, Pan, Garcia-Quintero, Huterer, Valogiannis, Ureña-López, Kamble, Parkinson, Kim, Zhao, Cervantes-Cota, Rohlf, Lozano-Rodríguez, Román-Herrera, Abdul-Karim, Aguilar, Ahlen, Alves, Andrade, Armengaud, Aviles, BenZvi, Bianchi, Brodzeller, Brooks, Burtin, Canning, Rosell, Casas, Castander, Charles, Chaussidon, Chaves-Montero, Chebat, Claybaugh, Cole, Cuceu, Dawson, de~la Macorra, de~Mattia, Deiosso, Demina, Dey, Dey, Ding, Doel, Eisenstein, Elbers, Ferraro, Font-Ribera, Forero-Romero, Garrison, Gaztañaga, Gil-Marín, Gontcho, Gonzalez-Morales, Gutierrez, Guy, Hahn, Herbold, Herrera-Alcantar, Honscheid, Howlett, Juneau, Kehoe, Kirkby, Kisner, Kremin, Lahav, Lamman, Landriau, Guillou, Leauthaud, Levi, Li, Magneville, Manera, Martini, Meisner, Mena-Fernández, Miquel, Moustakas, Santos, Muñoz-Gutiérrez, Myers, Nadathur, Niz, Noriega, Paillas, Palanque-Delabrouille, Percival, Pieri, Poppett, Prada, Pérez-Fernández, Pérez-Ràfols,
  Ramírez-Pérez, Rashkovetskyi, Ravoux, Ross, Rossi, Ruhlmann-Kleider, Samushia, Sanchez, Schlegel, Schubnell, Seo, Sinigaglia, Sprayberry, Tan, Tarlé, Taylor, Turner, Vargas-Magaña, Walther, Weaver, Wolfson, Yèche, Zarrouk, Zhou and Zou}]{extended_desi}
\bibinfo{author}{Lodha, K.}, \bibinfo{author}{Calderon, R.}, \bibinfo{author}{Matthewson, W.L.}, \bibinfo{author}{Shafieloo, A.}, \bibinfo{author}{Ishak, M.}, \bibinfo{author}{Pan, J.}, \bibinfo{author}{Garcia-Quintero, C.}, \bibinfo{author}{Huterer, D.}, \bibinfo{author}{Valogiannis, G.}, \bibinfo{author}{Ureña-López, L.A.}, \bibinfo{author}{Kamble, N.V.}, \bibinfo{author}{Parkinson, D.}, \bibinfo{author}{Kim, A.G.}, \bibinfo{author}{Zhao, G.B.}, \bibinfo{author}{Cervantes-Cota, J.L.}, \bibinfo{author}{Rohlf, J.}, \bibinfo{author}{Lozano-Rodríguez, F.}, \bibinfo{author}{Román-Herrera, J.O.}, \bibinfo{author}{Abdul-Karim, M.}, \bibinfo{author}{Aguilar, J.}, \bibinfo{author}{Ahlen, S.}, \bibinfo{author}{Alves, O.}, \bibinfo{author}{Andrade, U.}, \bibinfo{author}{Armengaud, E.}, \bibinfo{author}{Aviles, A.}, \bibinfo{author}{BenZvi, S.}, \bibinfo{author}{Bianchi, D.}, \bibinfo{author}{Brodzeller, A.}, \bibinfo{author}{Brooks, D.}, \bibinfo{author}{Burtin, E.}, \bibinfo{author}{Canning, R.},
  \bibinfo{author}{Rosell, A.C.}, \bibinfo{author}{Casas, L.}, \bibinfo{author}{Castander, F.J.}, \bibinfo{author}{Charles, M.}, \bibinfo{author}{Chaussidon, E.}, \bibinfo{author}{Chaves-Montero, J.}, \bibinfo{author}{Chebat, D.}, \bibinfo{author}{Claybaugh, T.}, \bibinfo{author}{Cole, S.}, \bibinfo{author}{Cuceu, A.}, \bibinfo{author}{Dawson, K.S.}, \bibinfo{author}{de~la Macorra, A.}, \bibinfo{author}{de~Mattia, A.}, \bibinfo{author}{Deiosso, N.}, \bibinfo{author}{Demina, R.}, \bibinfo{author}{Dey, A.}, \bibinfo{author}{Dey, B.}, \bibinfo{author}{Ding, Z.}, \bibinfo{author}{Doel, P.}, \bibinfo{author}{Eisenstein, D.J.}, \bibinfo{author}{Elbers, W.}, \bibinfo{author}{Ferraro, S.}, \bibinfo{author}{Font-Ribera, A.}, \bibinfo{author}{Forero-Romero, J.E.}, \bibinfo{author}{Garrison, L.H.}, \bibinfo{author}{Gaztañaga, E.}, \bibinfo{author}{Gil-Marín, H.}, \bibinfo{author}{Gontcho, S.G.A.}, \bibinfo{author}{Gonzalez-Morales, A.X.}, \bibinfo{author}{Gutierrez, G.}, \bibinfo{author}{Guy, J.},
  \bibinfo{author}{Hahn, C.}, \bibinfo{author}{Herbold, M.}, \bibinfo{author}{Herrera-Alcantar, H.K.}, \bibinfo{author}{Honscheid, K.}, \bibinfo{author}{Howlett, C.}, \bibinfo{author}{Juneau, S.}, \bibinfo{author}{Kehoe, R.}, \bibinfo{author}{Kirkby, D.}, \bibinfo{author}{Kisner, T.}, \bibinfo{author}{Kremin, A.}, \bibinfo{author}{Lahav, O.}, \bibinfo{author}{Lamman, C.}, \bibinfo{author}{Landriau, M.}, \bibinfo{author}{Guillou, L.L.}, \bibinfo{author}{Leauthaud, A.}, \bibinfo{author}{Levi, M.E.}, \bibinfo{author}{Li, Q.}, \bibinfo{author}{Magneville, C.}, \bibinfo{author}{Manera, M.}, \bibinfo{author}{Martini, P.}, \bibinfo{author}{Meisner, A.}, \bibinfo{author}{Mena-Fernández, J.}, \bibinfo{author}{Miquel, R.}, \bibinfo{author}{Moustakas, J.}, \bibinfo{author}{Santos, D.M.}, \bibinfo{author}{Muñoz-Gutiérrez, A.}, \bibinfo{author}{Myers, A.D.}, \bibinfo{author}{Nadathur, S.}, \bibinfo{author}{Niz, G.}, \bibinfo{author}{Noriega, H.E.}, \bibinfo{author}{Paillas, E.}, \bibinfo{author}{Palanque-Delabrouille,
  N.}, \bibinfo{author}{Percival, W.J.}, \bibinfo{author}{Pieri, M.M.}, \bibinfo{author}{Poppett, C.}, \bibinfo{author}{Prada, F.}, \bibinfo{author}{Pérez-Fernández, A.}, \bibinfo{author}{Pérez-Ràfols, I.}, \bibinfo{author}{Ramírez-Pérez, C.}, \bibinfo{author}{Rashkovetskyi, M.}, \bibinfo{author}{Ravoux, C.}, \bibinfo{author}{Ross, A.J.}, \bibinfo{author}{Rossi, G.}, \bibinfo{author}{Ruhlmann-Kleider, V.}, \bibinfo{author}{Samushia, L.}, \bibinfo{author}{Sanchez, E.}, \bibinfo{author}{Schlegel, D.}, \bibinfo{author}{Schubnell, M.}, \bibinfo{author}{Seo, H.}, \bibinfo{author}{Sinigaglia, F.}, \bibinfo{author}{Sprayberry, D.}, \bibinfo{author}{Tan, T.}, \bibinfo{author}{Tarlé, G.}, \bibinfo{author}{Taylor, P.}, \bibinfo{author}{Turner, W.}, \bibinfo{author}{Vargas-Magaña, M.}, \bibinfo{author}{Walther, M.}, \bibinfo{author}{Weaver, B.A.}, \bibinfo{author}{Wolfson, M.}, \bibinfo{author}{Yèche, C.}, \bibinfo{author}{Zarrouk, P.}, \bibinfo{author}{Zhou, R.}, \bibinfo{author}{Zou, H.},
  \bibinfo{year}{2025}.
\newblock \bibinfo{title}{{Extended Dark Energy analysis using DESI DR2 BAO measurements}}.
\newblock \URLprefix \url{https://arxiv.org/abs/2503.14743}, \href{http://arxiv.org/abs/2503.14743}{{\tt arXiv:2503.14743}}.
\bibitem[{Montani(2001)}]{matcre_montani2001}
\bibinfo{author}{Montani, G.}, \bibinfo{year}{2001}.
\newblock \bibinfo{title}{{Influence of particle creation on flat and negative curved FLRW universes}}.
\newblock \bibinfo{journal}{Classical and Quantum Gravity} \bibinfo{volume}{18}, \bibinfo{pages}{193}.
\bibitem[{Montani et~al.(2009)Montani, Battisti, Benini and Imponente}]{Primordial}
\bibinfo{author}{Montani, G.}, \bibinfo{author}{Battisti, M.V.}, \bibinfo{author}{Benini, R.}, \bibinfo{author}{Imponente, G.}, \bibinfo{year}{2009}.
\newblock \bibinfo{title}{{Primordial cosmology}}.
\newblock \bibinfo{publisher}{World Scientific}, \bibinfo{address}{Singapore}.
\newblock \DOIprefix\doi{10.1142/7235}.
\bibitem[{Montani et~al.(2024a)Montani, Carlevaro and Dainotti}]{montani-carlevaro-dainotti2024}
\bibinfo{author}{Montani, G.}, \bibinfo{author}{Carlevaro, N.}, \bibinfo{author}{Dainotti, M.G.}, \bibinfo{year}{2024}a.
\newblock \bibinfo{title}{{Slow-rolling scalar dynamics as solution for the Hubble tension}}.
\newblock \bibinfo{journal}{Physics of the Dark Universe} \bibinfo{volume}{44}, \bibinfo{pages}{101486}.
\newblock \DOIprefix\doi{10.1016/j.dark.2024.101486}.
\bibitem[{Montani et~al.(2025a)Montani, Carlevaro and Dainotti}]{montani-carlevaro-dainotti2025}
\bibinfo{author}{Montani, G.}, \bibinfo{author}{Carlevaro, N.}, \bibinfo{author}{Dainotti, M.G.}, \bibinfo{year}{2025}a.
\newblock \bibinfo{title}{{Running Hubble constant: evolutionary Dark Energy}}.
\newblock \bibinfo{journal}{Physics of the Dark Universe} , \bibinfo{pages}{101847}.
\bibitem[{Montani et~al.(2024b)Montani, Carlevaro and De~Angelis}]{matcre_montanimary}
\bibinfo{author}{Montani, G.}, \bibinfo{author}{Carlevaro, N.}, \bibinfo{author}{De~Angelis, M.}, \bibinfo{year}{2024}b.
\newblock \bibinfo{title}{{Modified gravity in the presence of matter creation: Scenario for the late Universe}}.
\newblock \bibinfo{journal}{Entropy} \bibinfo{volume}{26}, \bibinfo{pages}{662}.
\bibitem[{Montani et~al.(2024c)Montani, Carlevaro, Escamilla and Valentino}]{montani_hubbletension_4}
\bibinfo{author}{Montani, G.}, \bibinfo{author}{Carlevaro, N.}, \bibinfo{author}{Escamilla, L.A.}, \bibinfo{author}{Valentino, E.D.}, \bibinfo{year}{2024}c.
\newblock \bibinfo{title}{{Kinetic Model for Dark Energy -- Dark Matter Interaction: Scenario for the Hubble Tension}}.
\newblock \href{http://arxiv.org/abs/2404.15977}{{\tt arXiv:2404.15977}}.
\bibitem[{Montani et~al.(2023)Montani, De~Angelis, Bombacigno and Carlevaro}]{montani_hubbletension_5}
\bibinfo{author}{Montani, G.}, \bibinfo{author}{De~Angelis, M.}, \bibinfo{author}{Bombacigno, F.}, \bibinfo{author}{Carlevaro, N.}, \bibinfo{year}{2023}.
\newblock \bibinfo{title}{{Metric f(R) gravity with dynamical dark energy as a scenario for the Hubble tension}}.
\newblock \bibinfo{journal}{Monthly Notices of the Royal Astronomical Society: Letters} \bibinfo{volume}{527}, \bibinfo{pages}{L156–L161}.
\newblock \DOIprefix\doi{10.1093/mnrasl/slad159}.
\bibitem[{Montani et~al.(2025b)Montani, Fazzari, Carlevaro and Dainotti}]{montani2025_entropy}
\bibinfo{author}{Montani, G.}, \bibinfo{author}{Fazzari, E.}, \bibinfo{author}{Carlevaro, N.}, \bibinfo{author}{Dainotti, M.G.}, \bibinfo{year}{2025}b.
\newblock \bibinfo{title}{Two dynamical scenarios for the binned master sample interpretation}.
\newblock \URLprefix \url{https://arxiv.org/abs/2507.14048}, \href{http://arxiv.org/abs/2507.14048}{{\tt arXiv:2507.14048}}.
\bibitem[{Moresco(2015)}]{moresco_2015}
\bibinfo{author}{Moresco, M.}, \bibinfo{year}{2015}.
\newblock \bibinfo{title}{{Raising the bar: new constraints on the Hubble parameter with cosmic chronometers at $z\sim2$}}.
\newblock \bibinfo{journal}{Monthly Notices of the Royal Astronomical Society: Letters} \bibinfo{volume}{450}, \bibinfo{pages}{L16–L20}.
\newblock \URLprefix \url{http://dx.doi.org/10.1093/mnrasl/slv037}, \DOIprefix\doi{10.1093/mnrasl/slv037}.
\bibitem[{Moresco et~al.(2012)Moresco, Cimatti, Jimenez, Pozzetti, Zamorani, Bolzonella, Dunlop, Lamareille, Mignoli, Pearce, Rosati, Stern, Verde, Zucca, Carollo, Contini, Kneib, Fèvre, Lilly, Mainieri, Renzini, Scodeggio, Balestra, Gobat, McLure, Bardelli, Bongiorno, Caputi, Cucciati, de~la Torre, de~Ravel, Franzetti, Garilli, Iovino, Kampczyk, Knobel, Kovač, Borgne, Brun, Maier, Pelló, Peng, Perez-Montero, Presotto, Silverman, Tanaka, Tasca, Tresse, Vergani, Almaini, Barnes, Bordoloi, Bradshaw, Cappi, Chuter, Cirasuolo, Coppa, Diener, Foucaud, Hartley, Kamionkowski, Koekemoer, López-Sanjuan, McCracken, Nair, Oesch, Stanford and Welikala}]{Moresco_2012}
\bibinfo{author}{Moresco, M.}, \bibinfo{author}{Cimatti, A.}, \bibinfo{author}{Jimenez, R.}, \bibinfo{author}{Pozzetti, L.}, \bibinfo{author}{Zamorani, G.}, \bibinfo{author}{Bolzonella, M.}, \bibinfo{author}{Dunlop, J.}, \bibinfo{author}{Lamareille, F.}, \bibinfo{author}{Mignoli, M.}, \bibinfo{author}{Pearce, H.}, \bibinfo{author}{Rosati, P.}, \bibinfo{author}{Stern, D.}, \bibinfo{author}{Verde, L.}, \bibinfo{author}{Zucca, E.}, \bibinfo{author}{Carollo, C.}, \bibinfo{author}{Contini, T.}, \bibinfo{author}{Kneib, J.P.}, \bibinfo{author}{Fèvre, O.L.}, \bibinfo{author}{Lilly, S.}, \bibinfo{author}{Mainieri, V.}, \bibinfo{author}{Renzini, A.}, \bibinfo{author}{Scodeggio, M.}, \bibinfo{author}{Balestra, I.}, \bibinfo{author}{Gobat, R.}, \bibinfo{author}{McLure, R.}, \bibinfo{author}{Bardelli, S.}, \bibinfo{author}{Bongiorno, A.}, \bibinfo{author}{Caputi, K.}, \bibinfo{author}{Cucciati, O.}, \bibinfo{author}{de~la Torre, S.}, \bibinfo{author}{de~Ravel, L.}, \bibinfo{author}{Franzetti, P.},
  \bibinfo{author}{Garilli, B.}, \bibinfo{author}{Iovino, A.}, \bibinfo{author}{Kampczyk, P.}, \bibinfo{author}{Knobel, C.}, \bibinfo{author}{Kovač, K.}, \bibinfo{author}{Borgne, J.F.L.}, \bibinfo{author}{Brun, V.L.}, \bibinfo{author}{Maier, C.}, \bibinfo{author}{Pelló, R.}, \bibinfo{author}{Peng, Y.}, \bibinfo{author}{Perez-Montero, E.}, \bibinfo{author}{Presotto, V.}, \bibinfo{author}{Silverman, J.}, \bibinfo{author}{Tanaka, M.}, \bibinfo{author}{Tasca, L.}, \bibinfo{author}{Tresse, L.}, \bibinfo{author}{Vergani, D.}, \bibinfo{author}{Almaini, O.}, \bibinfo{author}{Barnes, L.}, \bibinfo{author}{Bordoloi, R.}, \bibinfo{author}{Bradshaw, E.}, \bibinfo{author}{Cappi, A.}, \bibinfo{author}{Chuter, R.}, \bibinfo{author}{Cirasuolo, M.}, \bibinfo{author}{Coppa, G.}, \bibinfo{author}{Diener, C.}, \bibinfo{author}{Foucaud, S.}, \bibinfo{author}{Hartley, W.}, \bibinfo{author}{Kamionkowski, M.}, \bibinfo{author}{Koekemoer, A.}, \bibinfo{author}{López-Sanjuan, C.}, \bibinfo{author}{McCracken, H.},
  \bibinfo{author}{Nair, P.}, \bibinfo{author}{Oesch, P.}, \bibinfo{author}{Stanford, A.}, \bibinfo{author}{Welikala, N.}, \bibinfo{year}{2012}.
\newblock \bibinfo{title}{{Improved constraints on the expansion rate of the Universe up to $z \sim 1.1$ from the spectroscopic evolution of cosmic chronometers}}.
\newblock \bibinfo{journal}{Journal of Cosmology and Astroparticle Physics} \bibinfo{volume}{2012}, \bibinfo{pages}{006–006}.
\newblock \URLprefix \url{http://dx.doi.org/10.1088/1475-7516/2012/08/006}, \DOIprefix\doi{10.1088/1475-7516/2012/08/006}.
\bibitem[{Moresco et~al.(2016)Moresco, Pozzetti, Cimatti, Jimenez, Maraston, Verde, Thomas, Citro, Tojeiro and Wilkinson}]{Moresco_2016}
\bibinfo{author}{Moresco, M.}, \bibinfo{author}{Pozzetti, L.}, \bibinfo{author}{Cimatti, A.}, \bibinfo{author}{Jimenez, R.}, \bibinfo{author}{Maraston, C.}, \bibinfo{author}{Verde, L.}, \bibinfo{author}{Thomas, D.}, \bibinfo{author}{Citro, A.}, \bibinfo{author}{Tojeiro, R.}, \bibinfo{author}{Wilkinson, D.}, \bibinfo{year}{2016}.
\newblock \bibinfo{title}{{A $6\%$ measurement of the Hubble parameter at $z\sim0.45$: direct evidence of the epoch of cosmic re-acceleration}}.
\newblock \bibinfo{journal}{Journal of Cosmology and Astroparticle Physics} \bibinfo{volume}{2016}, \bibinfo{pages}{014–014}.
\newblock \URLprefix \url{http://dx.doi.org/10.1088/1475-7516/2016/05/014}, \DOIprefix\doi{10.1088/1475-7516/2016/05/014}.
\bibitem[{Pedrotti et~al.(2025)Pedrotti, Jiang, Escamilla, da~Costa and Vagnozzi}]{Pedrotti:2024kpn}
\bibinfo{author}{Pedrotti, D.}, \bibinfo{author}{Jiang, J.Q.}, \bibinfo{author}{Escamilla, L.A.}, \bibinfo{author}{da~Costa, S.S.}, \bibinfo{author}{Vagnozzi, S.}, \bibinfo{year}{2025}.
\newblock \bibinfo{title}{{Multidimensionality of the Hubble tension: The roles of {\ensuremath{\Omega}}m and {\ensuremath{\omega}}c}}.
\newblock \bibinfo{journal}{Phys. Rev. D} \bibinfo{volume}{111}, \bibinfo{pages}{023506}.
\newblock \DOIprefix\doi{10.1103/PhysRevD.111.023506}, \href{http://arxiv.org/abs/2408.04530}{{\tt arXiv:2408.04530}}.
\bibitem[{Riess et~al.(1998)Riess, Filippenko, Challis, Clocchiatti, Diercks, Garnavich, Gilliland, Hogan, Jha, Kirshner et~al.}]{riess1998}
\bibinfo{author}{Riess, A.G.}, \bibinfo{author}{Filippenko, A.V.}, \bibinfo{author}{Challis, P.}, \bibinfo{author}{Clocchiatti, A.}, \bibinfo{author}{Diercks, A.}, \bibinfo{author}{Garnavich, P.M.}, \bibinfo{author}{Gilliland, R.L.}, \bibinfo{author}{Hogan, C.J.}, \bibinfo{author}{Jha, S.}, \bibinfo{author}{Kirshner, R.P.}, et~al., \bibinfo{year}{1998}.
\newblock \bibinfo{title}{{Observational evidence from supernovae for an accelerating universe and a cosmological constant}}.
\newblock \bibinfo{journal}{The astronomical journal} \bibinfo{volume}{116}, \bibinfo{pages}{1009}.
\bibitem[{Riess et~al.(2016)Riess, Macri, Hoffmann, Scolnic, Casertano, Filippenko, Tucker, Reid, Jones, Silverman et~al.}]{riess2016}
\bibinfo{author}{Riess, A.G.}, \bibinfo{author}{Macri, L.M.}, \bibinfo{author}{Hoffmann, S.L.}, \bibinfo{author}{Scolnic, D.}, \bibinfo{author}{Casertano, S.}, \bibinfo{author}{Filippenko, A.V.}, \bibinfo{author}{Tucker, B.E.}, \bibinfo{author}{Reid, M.J.}, \bibinfo{author}{Jones, D.O.}, \bibinfo{author}{Silverman, J.M.}, et~al., \bibinfo{year}{2016}.
\newblock \bibinfo{title}{{A $2.4\%$ determination of the local value of the Hubble constant}}.
\newblock \bibinfo{journal}{The Astrophysical Journal} \bibinfo{volume}{826}, \bibinfo{pages}{56}.
\bibitem[{Riess et~al.(2022)Riess, Yuan, Macri, Scolnic, Brout, Casertano, Jones, Murakami, Anand, Breuval et~al.}]{SH0ES}
\bibinfo{author}{Riess, A.G.}, \bibinfo{author}{Yuan, W.}, \bibinfo{author}{Macri, L.M.}, \bibinfo{author}{Scolnic, D.}, \bibinfo{author}{Brout, D.}, \bibinfo{author}{Casertano, S.}, \bibinfo{author}{Jones, D.O.}, \bibinfo{author}{Murakami, Y.}, \bibinfo{author}{Anand, G.S.}, \bibinfo{author}{Breuval, L.}, et~al., \bibinfo{year}{2022}.
\newblock \bibinfo{title}{{A comprehensive measurement of the local value of the Hubble constant with 1 km s$^{-1}$ Mpc${-1}$ uncertainty from the Hubble Space Telescope and the SH0ES team}}.
\newblock \bibinfo{journal}{The Astrophysical journal letters} \bibinfo{volume}{934}, \bibinfo{pages}{L7}.
\newblock \DOIprefix\doi{10.3847/2041-8213/ac5c5b}.
\bibitem[{Sahni et~al.(2008)Sahni, Shafieloo and Starobinsky}]{Omz_starobinsky}
\bibinfo{author}{Sahni, V.}, \bibinfo{author}{Shafieloo, A.}, \bibinfo{author}{Starobinsky, A.A.}, \bibinfo{year}{2008}.
\newblock \bibinfo{title}{{Two new diagnostics of dark energy}}.
\newblock \bibinfo{journal}{Phys. Rev. D} \bibinfo{volume}{78}, \bibinfo{pages}{103502}.
\newblock \DOIprefix\doi{10.1103/PhysRevD.78.103502}, \href{http://arxiv.org/abs/0807.3548}{{\tt arXiv:0807.3548}}.
\bibitem[{Schiavone and Montani(2025)}]{schiavone2024}
\bibinfo{author}{Schiavone, T.}, \bibinfo{author}{Montani, G.}, \bibinfo{year}{2025}.
\newblock \bibinfo{title}{{Evolution of an effective Hubble constant in f (R) modified gravity}}.
\newblock \bibinfo{journal}{Nuovo Cim. C} \bibinfo{volume}{48}, \bibinfo{pages}{105}.
\newblock \DOIprefix\doi{10.1393/ncc/i2025-25105-3}, \href{http://arxiv.org/abs/2408.01410}{{\tt arXiv:2408.01410}}.
\bibitem[{Schiavone et~al.(2023)Schiavone, Montani and Bombacigno}]{schiavone2023}
\bibinfo{author}{Schiavone, T.}, \bibinfo{author}{Montani, G.}, \bibinfo{author}{Bombacigno, F.}, \bibinfo{year}{2023}.
\newblock \bibinfo{title}{{f(R) gravity in the Jordan frame as a paradigm for the Hubble tension}}.
\newblock \bibinfo{journal}{Monthly Notices of the Royal Astronomical Society: Letters} \bibinfo{volume}{522}, \bibinfo{pages}{L72–L77}.
\newblock \DOIprefix\doi{10.1093/mnrasl/slad041}.
\bibitem[{Schiavone et~al.(2022)Schiavone, Montani, Dainotti, De~Simone, Rinaldi and Lambiase}]{schiavone2022running}
\bibinfo{author}{Schiavone, T.}, \bibinfo{author}{Montani, G.}, \bibinfo{author}{Dainotti, M.G.}, \bibinfo{author}{De~Simone, B.}, \bibinfo{author}{Rinaldi, E.}, \bibinfo{author}{Lambiase, G.}, \bibinfo{year}{2022}.
\newblock \bibinfo{title}{{Running Hubble constant from the SNe Ia Pantheon sample?}}
\newblock \bibinfo{journal}{arXiv preprint arXiv:2205.07033} .
\bibitem[{Scolnic et~al.(2022)Scolnic, Brout, Carr, Riess, Davis, Dwomoh, Jones, Ali, Charvu, Chen, Peterson, Popovic, Rose, Wood, Brown, Chambers, Coulter, Dettman, Dimitriadis, Filippenko, Foley, Jha, Kilpatrick, Kirshner, Pan, Rest, Rojas-Bravo, Siebert, Stahl and Zheng}]{Scolnic_2022}
\bibinfo{author}{Scolnic, D.}, \bibinfo{author}{Brout, D.}, \bibinfo{author}{Carr, A.}, \bibinfo{author}{Riess, A.G.}, \bibinfo{author}{Davis, T.M.}, \bibinfo{author}{Dwomoh, A.}, \bibinfo{author}{Jones, D.O.}, \bibinfo{author}{Ali, N.}, \bibinfo{author}{Charvu, P.}, \bibinfo{author}{Chen, R.}, \bibinfo{author}{Peterson, E.R.}, \bibinfo{author}{Popovic, B.}, \bibinfo{author}{Rose, B.M.}, \bibinfo{author}{Wood, C.M.}, \bibinfo{author}{Brown, P.J.}, \bibinfo{author}{Chambers, K.}, \bibinfo{author}{Coulter, D.A.}, \bibinfo{author}{Dettman, K.G.}, \bibinfo{author}{Dimitriadis, G.}, \bibinfo{author}{Filippenko, A.V.}, \bibinfo{author}{Foley, R.J.}, \bibinfo{author}{Jha, S.W.}, \bibinfo{author}{Kilpatrick, C.D.}, \bibinfo{author}{Kirshner, R.P.}, \bibinfo{author}{Pan, Y.C.}, \bibinfo{author}{Rest, A.}, \bibinfo{author}{Rojas-Bravo, C.}, \bibinfo{author}{Siebert, M.R.}, \bibinfo{author}{Stahl, B.E.}, \bibinfo{author}{Zheng, W.}, \bibinfo{year}{2022}.
\newblock \bibinfo{title}{{The Pantheon+ Analysis: The Full Data Set and Light-curve Release}}.
\newblock \bibinfo{journal}{The Astrophysical Journal} \bibinfo{volume}{938}, \bibinfo{pages}{113}.
\newblock \DOIprefix\doi{10.3847/1538-4357/ac8b7a}.
\bibitem[{Scolnic et~al.(2018)Scolnic, Jones, Rest, Pan, Chornock, Foley, Huber, Kessler, Narayan, Riess et~al.}]{scolnic2018}
\bibinfo{author}{Scolnic, D.M.}, \bibinfo{author}{Jones, D.}, \bibinfo{author}{Rest, A.}, \bibinfo{author}{Pan, Y.}, \bibinfo{author}{Chornock, R.}, \bibinfo{author}{Foley, R.}, \bibinfo{author}{Huber, M.}, \bibinfo{author}{Kessler, R.}, \bibinfo{author}{Narayan, G.}, \bibinfo{author}{Riess, A.}, et~al., \bibinfo{year}{2018}.
\newblock \bibinfo{title}{{The complete light-curve sample of spectroscopically confirmed SNe Ia from Pan-STARRS1 and cosmological constraints from the combined pantheon sample}}.
\newblock \bibinfo{journal}{The Astrophysical Journal} \bibinfo{volume}{859}, \bibinfo{pages}{101}.
\bibitem[{Torrado and Lewis(2021)}]{cobaya}
\bibinfo{author}{Torrado, J.}, \bibinfo{author}{Lewis, A.}, \bibinfo{year}{2021}.
\newblock \bibinfo{title}{{Cobaya: Code for Bayesian Analysis of hierarchical physical models}}.
\newblock \bibinfo{journal}{Journal of Cosmology and Astroparticle Physics} \bibinfo{volume}{2021}, \bibinfo{pages}{057}.
\bibitem[{Trotta(2008)}]{Bayes_trotta}
\bibinfo{author}{Trotta, R.}, \bibinfo{year}{2008}.
\newblock \bibinfo{title}{{Bayes in the sky: Bayesian inference and model selection in cosmology}}.
\newblock \bibinfo{journal}{Contemporary Physics} \bibinfo{volume}{49}, \bibinfo{pages}{71--104}.
\bibitem[{Vagnozzi(2020)}]{Vagnozzi:2019ezj}
\bibinfo{author}{Vagnozzi, S.}, \bibinfo{year}{2020}.
\newblock \bibinfo{title}{{New physics in light of the $H_0$ tension: An alternative view}}.
\newblock \bibinfo{journal}{Phys. Rev. D} \bibinfo{volume}{102}, \bibinfo{pages}{023518}.
\newblock \DOIprefix\doi{10.1103/PhysRevD.102.023518}, \href{http://arxiv.org/abs/1907.07569}{{\tt arXiv:1907.07569}}.
\bibitem[{Vagnozzi(2023)}]{Vagnozzi:2023nrq}
\bibinfo{author}{Vagnozzi, S.}, \bibinfo{year}{2023}.
\newblock \bibinfo{title}{{Seven Hints That Early-Time New Physics Alone Is Not Sufficient to Solve the Hubble Tension}}.
\newblock \bibinfo{journal}{Universe} \bibinfo{volume}{9}, \bibinfo{pages}{393}.
\newblock \DOIprefix\doi{10.3390/universe9090393}, \href{http://arxiv.org/abs/2308.16628}{{\tt arXiv:2308.16628}}.
\bibitem[{Weinberg(1972)}]{weinberg-grav-cosm}
\bibinfo{author}{Weinberg, S.}, \bibinfo{year}{1972}.
\newblock \bibinfo{title}{{Gravitation and Cosmology: Principles and Applications of the General Theory of Relativity}}.
\newblock \bibinfo{publisher}{Wiley}.

\end{thebibliography}






\end{document}